\documentclass{article}
\usepackage{amsmath}
\usepackage{amsthm}
\usepackage{amsfonts}
\usepackage{amssymb}
\usepackage{fullpage}
\usepackage{mathtools}
\usepackage{color}
\usepackage{graphicx}
\graphicspath{{Figures/}}

\newcommand*{\myabstract}[1]{ \begin{abstract} #1 \end{abstract} }
\date{March 31, 2017}
\begin{document}
%%%%%%%%%%%%%%%%%%%%%
\title{\bf Sensitivity of Love and quasi-Rayleigh waves to model parameters}
\author{
D.R. Dalton%
\footnote{
Department of Earth Sciences, Memorial University of Newfoundland, Canada, {\tt dalton.nfld@gmail.com}}\,,
M.A. Slawinski%
\footnote{
Department of Earth Sciences, Memorial University of Newfoundland, Canada, {\tt mslawins@mac.com}}\,,
P. Stachura%
\footnote{
Faculty of Applied Informatics and Mathematics, SGGW, Poland, {\tt piotr\_stachura1@sggw.pl}}\,,
T. Stanoev%
\footnote{
Department of Earth Sciences, Memorial University of Newfoundland, Canada, {\tt theodore.stanoev@gmail.com}}
}
\maketitle
\myabstract{
We examine the sensitivity of the Love and the quasi-Rayleigh waves to model parameters.
Both waves are guided waves that propagate in the same model of an elastic layer above an elastic halfspace.  
We study their dispersion curves without any simplifying assumptions, beyond the standard approach of elasticity theory in isotropic media.
We examine the sensitivity of both waves to elasticity parameters, frequency and layer thickness, for varying frequency and different modes.
In the case of Love waves, we derive and plot the absolute value of a dimensionless sensitivity coefficient in terms of partial derivatives, and perform an analysis to find the optimum frequency for determining the layer thickness.
For a coherency of the background information, we briefly review the Love-wave dispersion relation and provide details of the less common derivation of the quasi-Rayleigh relation in an appendix. 
We compare that derivation to past results in the literature, finding certain discrepancies among them.}
%%%%%%%%%%%%%%%%%%%%%%%%%%%%
\section{Introduction}
%%%%%%%%%%%%%%%%%%%%%%%%%%%%
In this paper, we examine the sensitivity of the Love and quasi-Rayleigh waves to model parameters.
This study provides insight into the reliability of inferences of model properties from data.
Herein, sensitivity refers to relations between the wave properties and model parameters in the context of dispersion relations.
It does not refer to a misfit between a model and experimental data.

Both waves propagate within an elastic isotropic layer over an elastic isotropic halfspace.
This provides a redundancy of information, since both waves are described in terms of the same model parameters.
To examine the sensitivity---given the elasticity parameters, mass densities and the thickness of the layer, as well as the frequency of the signal---we study expressions for the speeds of the waves that correspond to different modes for either wave.
In general, each wave has an infinite number of modes and each mode propagates with a different speed.
However, for a given frequency, there is a finite number of modes, and hence, speeds.  

On the surface, the two waves exhibit displacements that are perpendicular to each other.
The displacements of the Love wave are in the horizontal plane and perpendicular to the direction of propagation.
The displacements of the quasi-Rayleigh wave are in the vertical plane, and---on the surface---are parallel to the direction of propagation.  
Thus, different speeds and displacement directions render these waves and their modes empirically distinguishable.

The quasi-Rayleigh wave shares many similarities with the classical Rayleigh wave, but is not restricted to the halfspace alone.
In literature, the quasi-Rayleigh wave has been also referred to as Rayleigh-type wave, Rayleigh-like wave, generalized Rayleigh wave, Rayleigh-Lamb wave and Rayleigh wave in inhomogeneous media.

Seismological information, such as wave speeds measured on the surface, allow us to infer properties of the subsurface.
Herein, to gain insight into such an inverse problem, we examine the sensitivity of the forward one.
Wave speeds corresponding to different modes of either wave exhibit different sensitivities to model parameters.

Motivated by the accuracy of modern seismic measurements and availability of computational tools, we study the Love and quasi-Rayleigh dispersion curves without any simplifying assumptions, beyond the standard approach of elasticity theory in isotropic media.

The presented concept of studying sensitivities exhibits certain similarities to the recent work of Lucena and Taioli~\cite{Lucena}, but with a different approach and scope.
The novelty of this study consists of an analysis of the sensitivity of the dispersion relation to elasticity parameters and layer thickness for both Love and quasi-Rayleigh waves for varying frequencies and different modes. 
Furthermore, we formulate a dimensionless sensitivity coefficient analysis from which we obtain the optimum frequency of Love waves for determining layer thickness.

We begin this paper by examining Love waves and proceed to quasi-Rayleigh waves.
In either case, we discuss the sensitivity of the dispersion relations to elasticity parameters and layer thickness for varying frequencies and different modes.
We highlight the sensitivity results in Section~\ref{sec:SensitivityResults} and suggest further research directions in Section~\ref{sec:FutureWork}.
This paper contains three appendices, which consist, respectively, of the derivation of the quasi-Rayleigh dispersion relation, calculations of its determinant and a comparison to literature results.
%%%%%%%%%%%%%%%%%%%%%%%%%%%%
\section{Love waves}
\label{sec:LovWav}
%%%%%%%%%%%%%%%%%%%%%%%%%%%%
\subsection{Material properties}
%%%%%%%%%%%%%%%%%%%%%%%%%%%%
We wish to examine the sensitivity of guided~$SH$ waves to model parameters.
For this purpose, we consider an elastic layer with mass density,~$\rho^u$\,, elasticity parameter,~$C_{44}^u$\,, and hence,~$S$-wave propagation speed,~\mbox{$\beta^u=\sqrt{C_{44}^u/\rho^u\strut}$}\,.
Also, we consider an elastic halfspace with~$\rho^d$\,,~$C_{44}^d$ and~\mbox{$\beta^d=\sqrt{C_{44}^d/\rho^d\strut}$}.
We set the coordinate system in such a manner that the surface is at~$x_3=0$ and the interface is at~$x_3=Z$\,, with the~$x_3$-axis positive downwards.

For details of the derivation of the Love wave dispersion relation, see, for instance, Slawinski~\cite[Chapter~6]{SlawinskiWS2016} and references therein.
The dispersion relation for Love waves can be written as
\begin{equation}
\label{eq:LoveDisp}
\tan\left(\omega\sqrt{\frac{1}{(\beta^u)^2}-\frac{1}{v_\ell^2}}\,Z\right)
= 
\frac{
C_{44}^d\,\sqrt{\dfrac{\mathstrut 1}{v_\ell^2}-\dfrac{1}{(\beta^d)^2}}
}{
C_{44}^u\,\sqrt{\mathstrut\dfrac{\mathstrut 1}{(\beta^u)^2}-\dfrac{1}{v_\ell^2}}
}\,.
\end{equation}
To plot expression~(\ref{eq:LoveDisp}), we rewrite it as an expression equal to zero, where
{%
\begin{equation}
\mathrm{D}_\ell=
2\left(C_{44}^u\sqrt{\left(\frac{v_\ell}{\beta^u}\right)^{\!\!2}-1}\,\sin\left(\frac{\omega Z}{\beta^u}\sqrt{1-\left(\frac{\beta^u}{v_\ell}\right)^{\!\!2}}\,\right)- 
C_{44}^d\sqrt{1- \left(\frac{v_\ell}{\beta^d}\right)^{\!\!2}}\,\cos\left(\frac{\omega Z}{\beta^u}\sqrt{1-\left(\frac{\beta^u}{v_\ell}\right)^{\!\!2}}\,\right)\right)=0\,.
\label{eq:lovedisp2}
\end{equation}}
We plot expression~(\ref{eq:lovedisp2}) within a model in which we set the layer thickness~$Z=500~{\rm m}$\,, the  two elasticity parameters, $C_{44}^u=0.88\times10^{10}~{\rm N}/{\rm m}^2$ and~$C_{44}^d=4.16\times10^{10}~{\rm N}/{\rm m}^2$\,, and mass densities, $\rho^u=2200~{\rm kg}/{\rm m}^3$ and~$\rho^d=2600~{\rm kg}/{\rm m}^3$\,.
It follows that the~$S$-wave propagation speeds in the layer and the halfspace are~$\beta^u=2000~{\rm m}/{\rm s}$ and~$\beta^d=4000~{\rm m}/{\rm s}$\,, respectively.
Such a model could represent a sandstone layer over a granite halfspace.

\begin{figure}
\begin{center}
\includegraphics[scale=0.5]{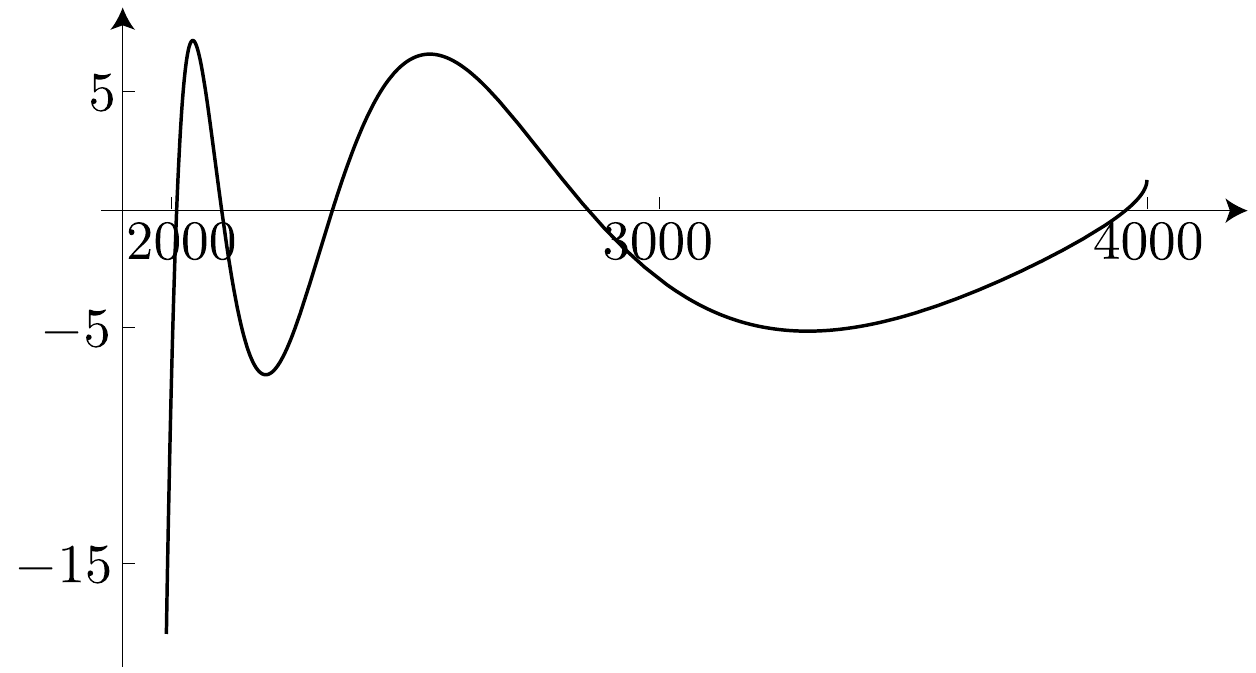}\hspace{0.5in}
\includegraphics[scale=0.5]{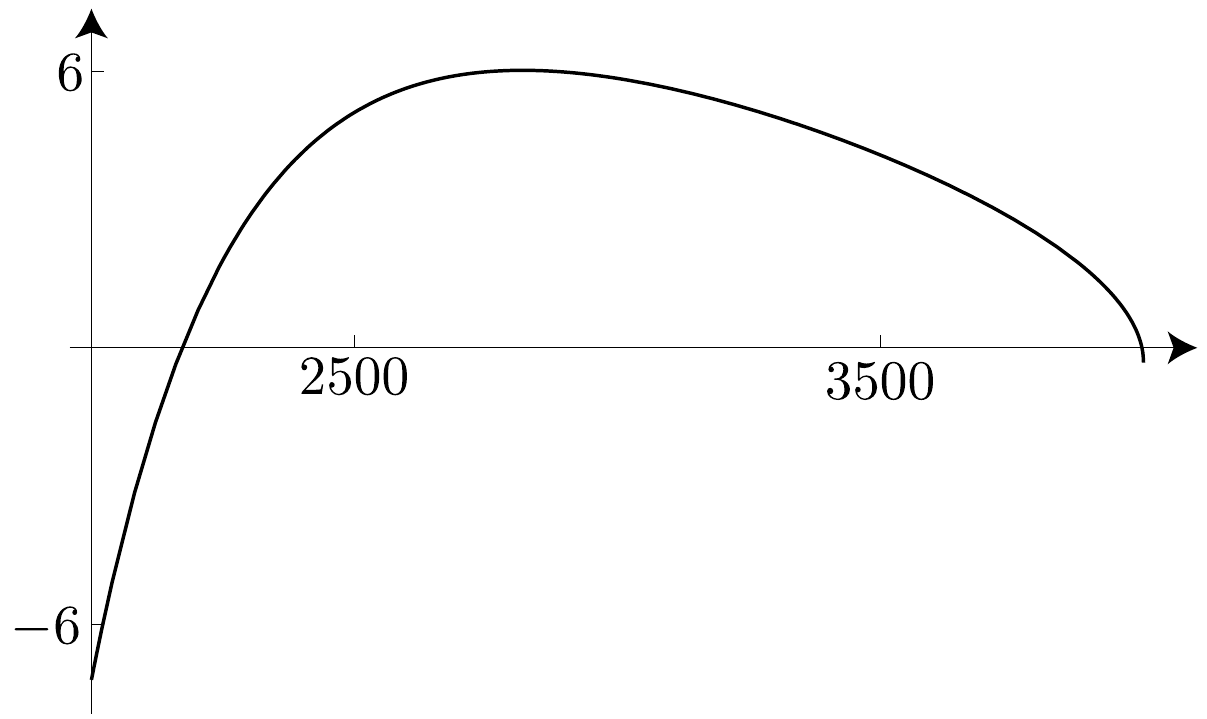}
%from love3x3.nb
\end{center}
\caption{\small{$\mathrm{D}_\ell$\,, defined in expression~(\ref{eq:lovedisp2}), as a function of speed,~$v_\ell$\,.
On the left, for~$\omega=60~{\rm s}^{-1}$\,, there are five roots:
${v_\ell=2010.7~{\rm m}/{\rm s}}$\,,
$v_\ell=2102.76~{\rm m}/{\rm s}$\,, 
$v_\ell=2330.44~{\rm m}/{\rm s}$\,,
$v_\ell=2853.13~{\rm m}/{\rm s}$ and
$v_\ell=3958.53~{\rm m}/{\rm s}$\,.
On the right, for~$\omega=15~{\rm s}^{-1}$\,,  
there are two roots:
$v_\ell=2172.48~{\rm m}/{\rm s}$ and~$v_\ell=3997.01~{\rm m}/{\rm s}$\,.
The values on the vertical axes are to be multiplied by~$10^{10}$\,.}}
\label{fig:love}
\end{figure}

\begin{figure}
\begin{center}
\includegraphics[scale=0.5]{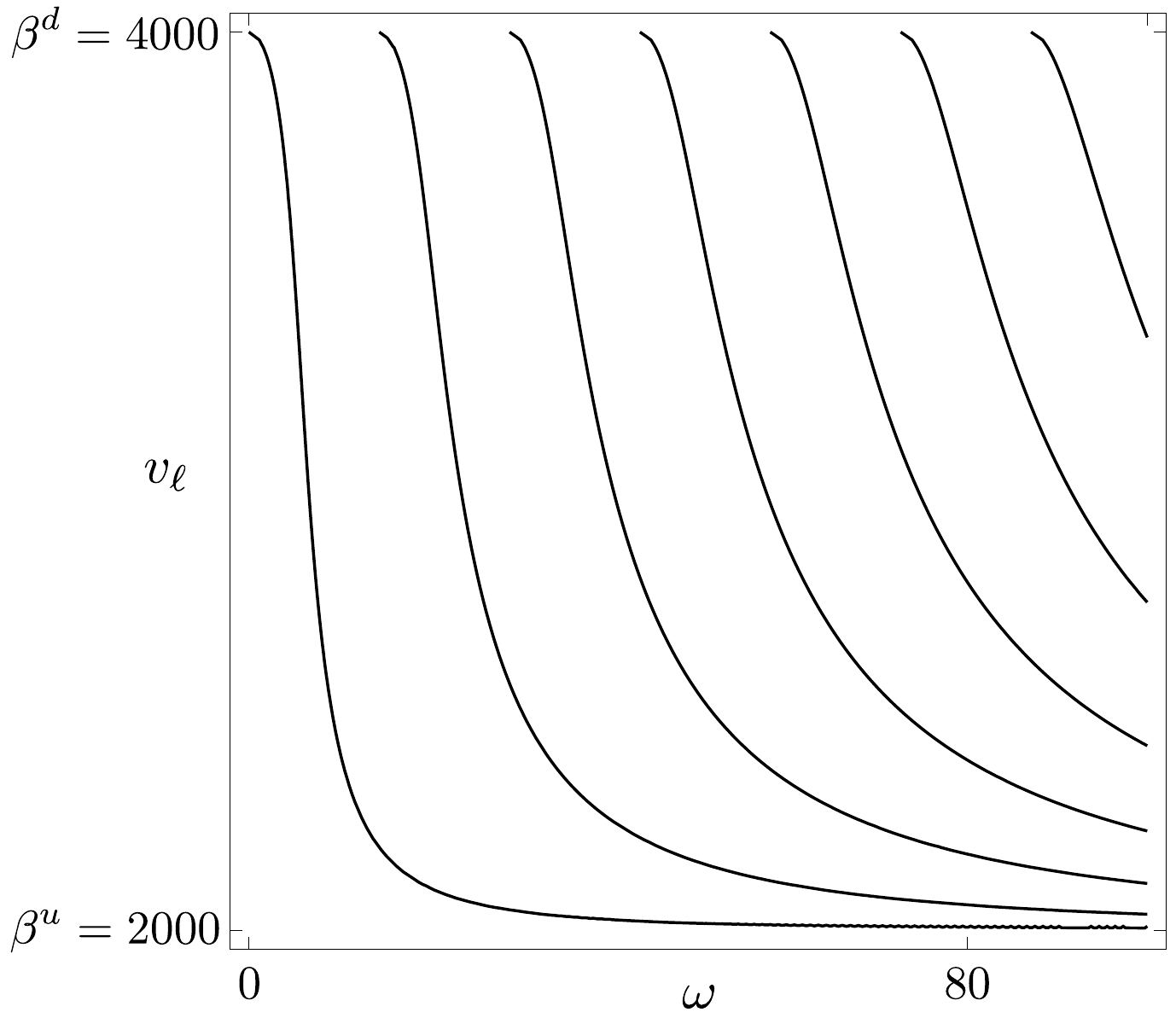}
\end{center}
\caption{\small{
The Love-wave dispersion curves,~$\mathrm{D}_\ell=0$\,, defined in expression~(\ref{eq:lovedisp2}), as a function of speed,~$v_\ell$\,, and frequency,~$\omega$\,.}}
%from love3x3contour-w5.nb
\label{fig:love3x3contourw5-2}
\end{figure}

Examining the left and right plots of Figure~\ref{fig:love}, as well as the dispersion curves in Figure~\ref{fig:love3x3contourw5-2}, we see that, for high frequency,~$\omega=60~{\rm s}^{-1}$\,, the speed of the fundamental mode of the Love wave,~$v_\ell=2011~{\rm m}/{\rm s}$\,, approaches~$\beta^u$\,, from above.
The values of~$v_\ell$ are computed numerically.
This result is in agreement with Ud\'{i}as~\cite[p.~196]{Udias1999}, and with the general theory of Love waves.%
%%%%%%%%%%%%%%%%%%%%%%%%%%%%
\subsection{Sensitivity of dispersion relation}
\label{sec:lovesens}
%%%%%%%%%%%%%%%%%%%%%%%%%%%%
We wish to examine effects of elasticity parameters and layer thickness on the dispersion relation, for various frequencies and different modes.
To do so, we examine effects of these quantities on the value of~$\mathrm{D}_\ell$\,, defined in expression~(\ref{eq:lovedisp2}).
Specifically, we examine~$\mathrm{D}_\ell$ as a function of~$C_{44}^d$ and~$C_{44}^u$\,, for two distinct frequencies and for fixed values of~$v_\ell$\,, which correspond to particular modes.

\begin{figure}
\begin{center}
\hspace*{-0.1in}\includegraphics[scale=0.5]{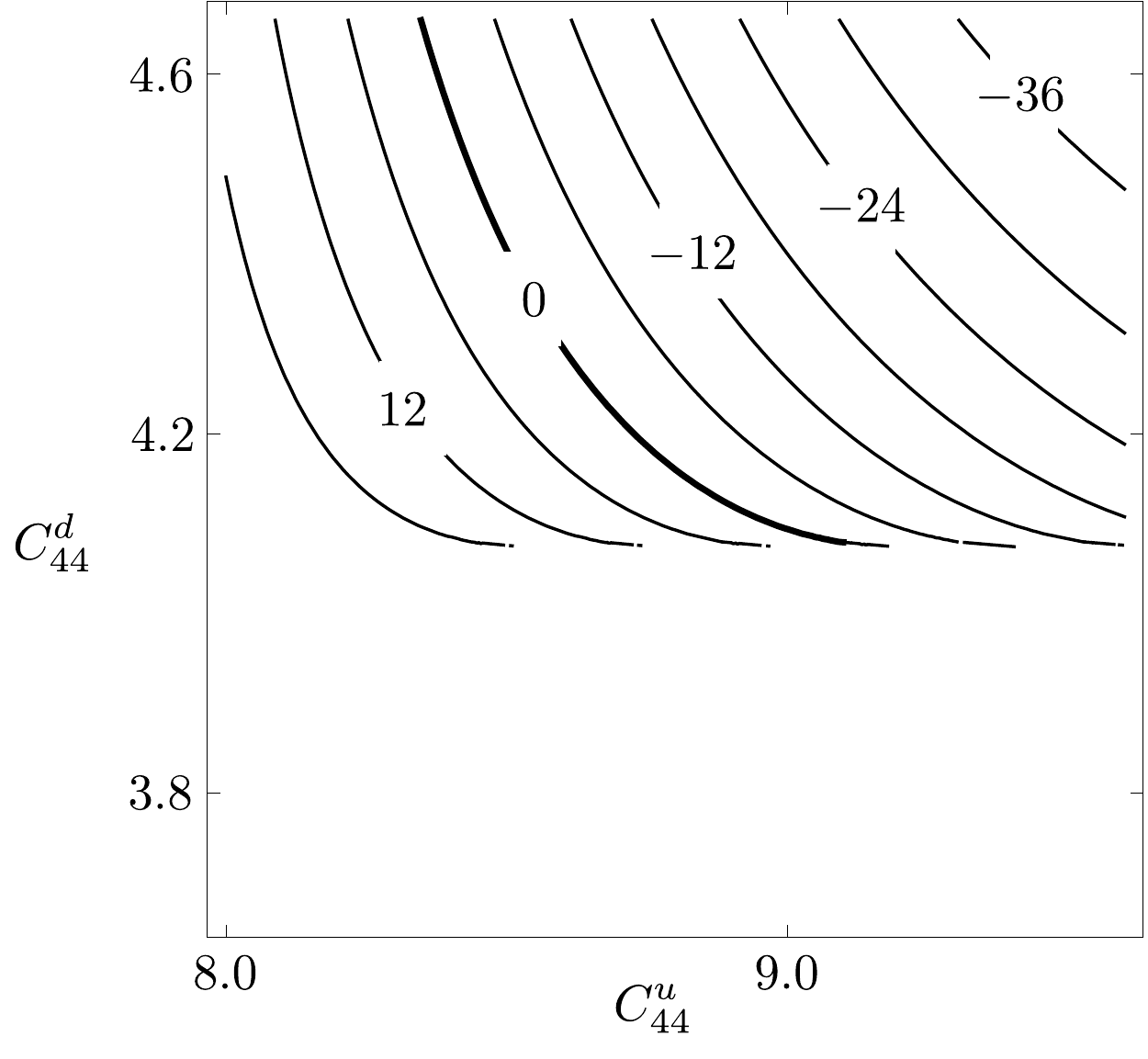}\hspace*{0.5in}\includegraphics[scale=0.5]{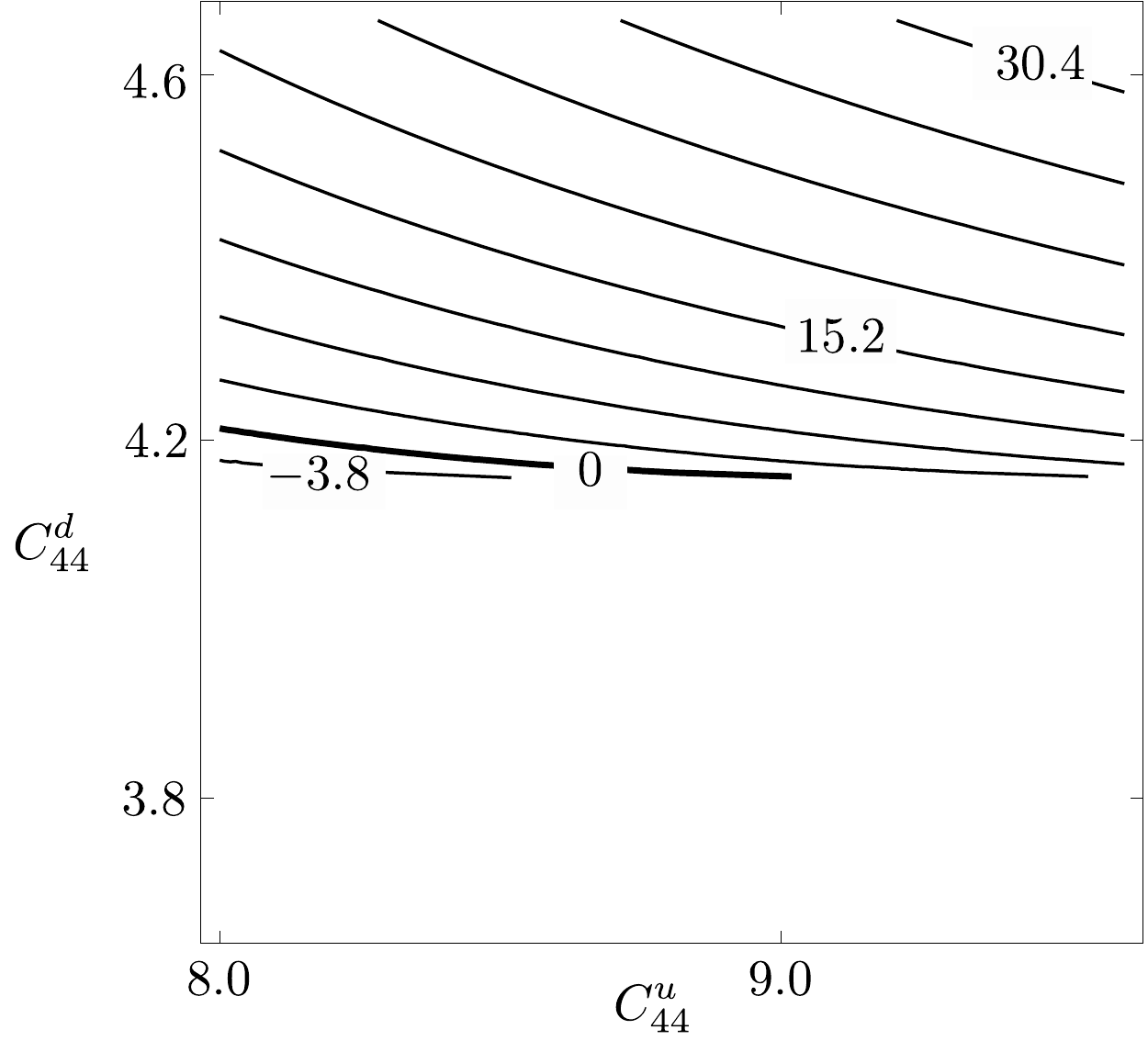}
%from love3x3contour-2.nb
\end{center}
\caption{\small{
$\mathrm{D}_\ell/10^{9}$\,, defined in expression~(\ref{eq:lovedisp2}),  as a function of the elasticity parameters,~$C_{44}^u$ and~$C_{44}^d$\,, 
for
$\omega=60~{\rm s}^{-1}$ and 
$v_\ell=3958.53~{\rm m}/{\rm s}$\,, on the left,
and for
$\omega=15~{\rm s}^{-1}$ and 
$v_\ell=3997.01~{\rm m}/{\rm s}$\,, on the right.
The values on the horizontal and vertical axes are to be multiplied by~$10^9$ and~$10^{10}$\,, respectively.  Both plots demonstrate sensitivity to
$C_{44}^u$ and~$C_{44}^d$ but the right plot, for lower frequency, demonstrates a greater sensitivity to~$C_{44}^d$\,.
}}
\label{fig:lc-vhigh}
\end{figure}

\begin{figure}
\begin{center}
\hspace*{-0.1in}\includegraphics[scale=0.5]{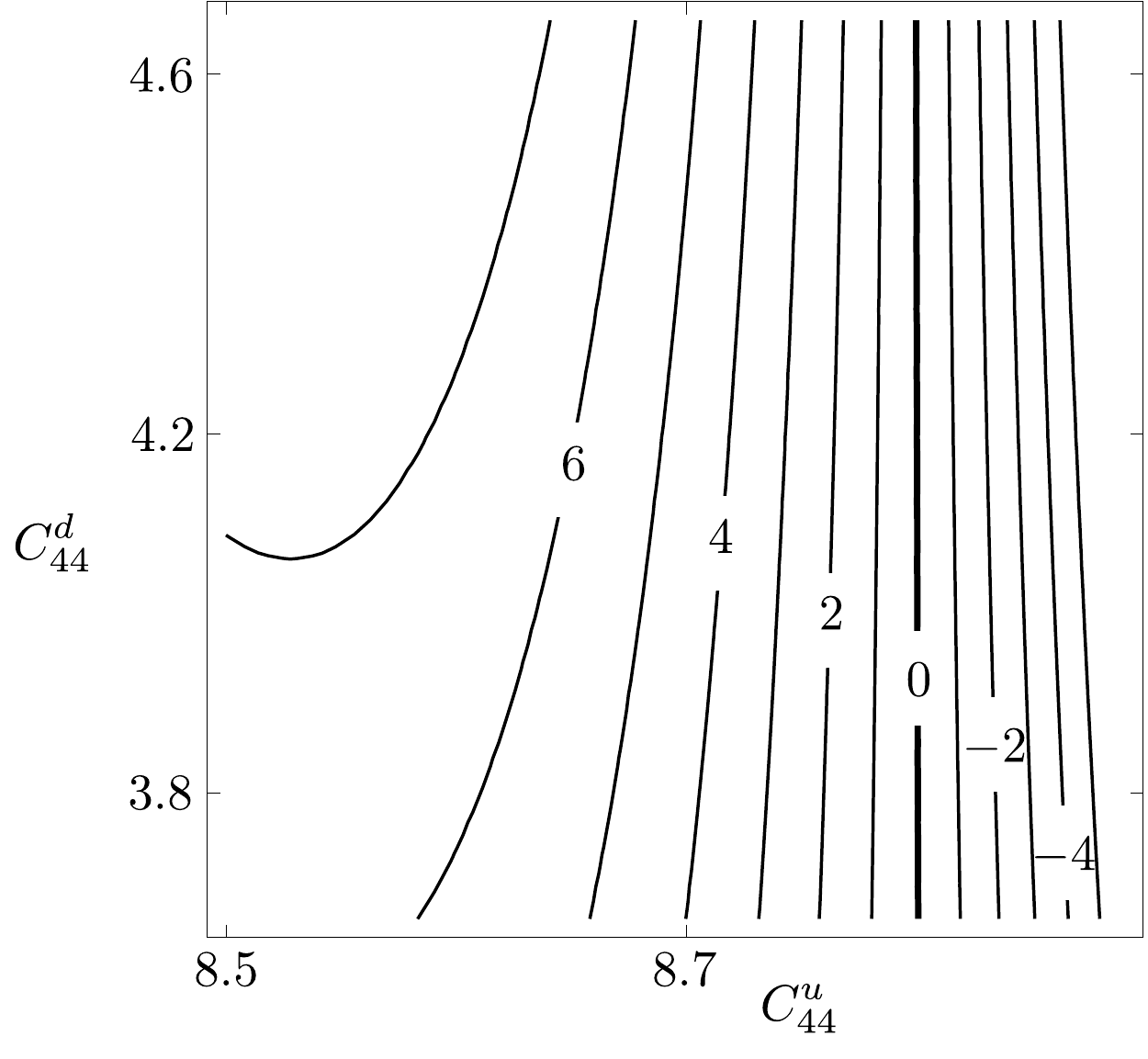}\hspace*{0.5in}\includegraphics[scale=0.5]{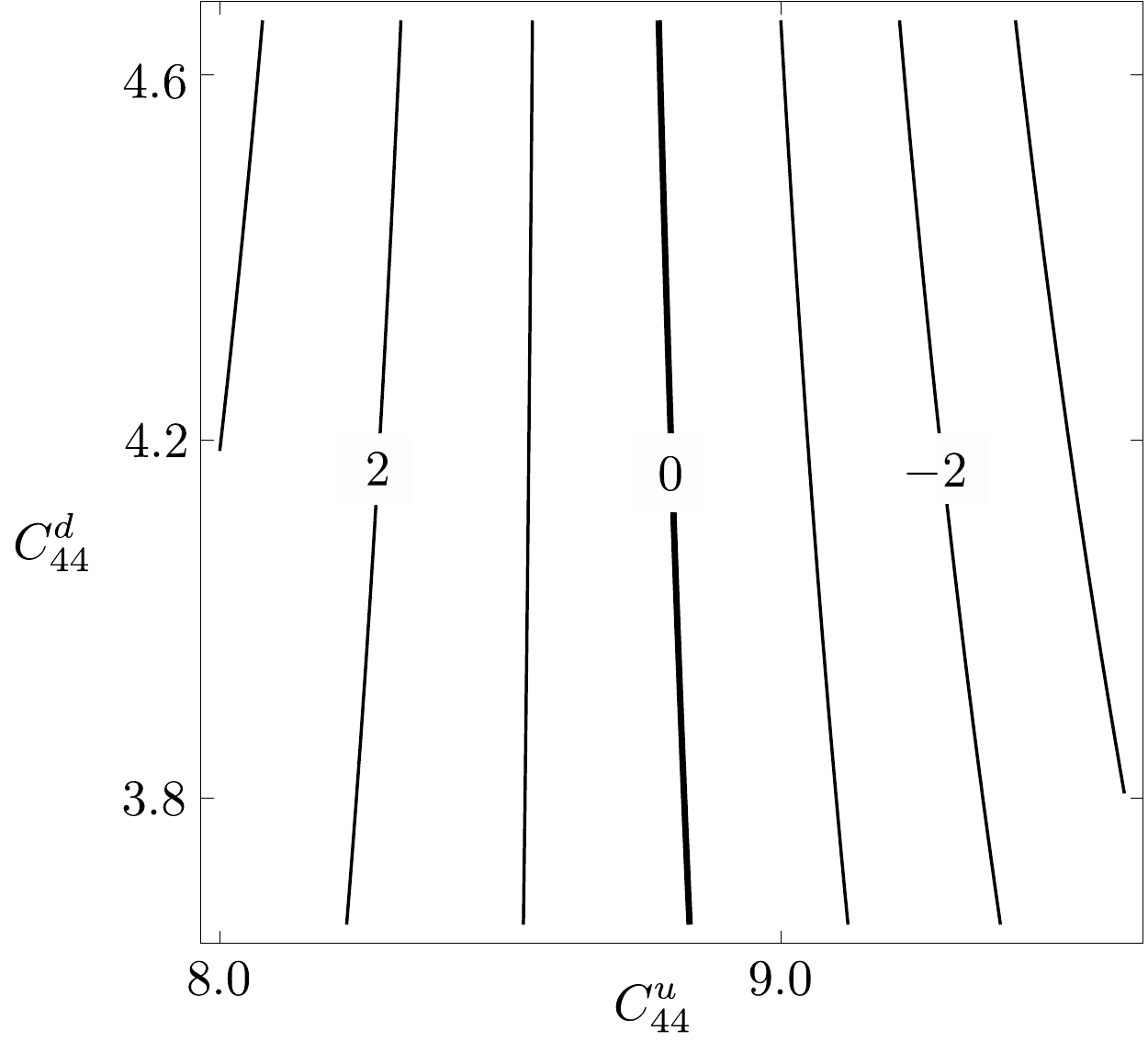}
%from love3x3contour-2.nb
\end{center}
\caption{\small{
$\mathrm{D}_\ell/10^{10}$\,, defined in expression~(\ref{eq:lovedisp2}),  as a function of the elasticity parameters,~$C_{44}^u$ and~$C_{44}^d$\,, where
$\omega=60~{\rm s}^{-1}$ and
$v_\ell=2010.7~{\rm m}/{\rm s}$\,,
and where
$\omega=15~{\rm s}^{-1}$ and
$v_\ell=2172.48~{\rm m}/{\rm s}$\,,
for left and right, respectively.
The values on the horizontal and vertical axes are to be multiplied by~$10^9$ and~$10^{10}$\,, respectively.
We see from the zero contours, which are near vertical, that there is sensitivity to~$C_{44}^u$ and not to~$C_{44}^d$\,, and this contrast is slightly more pronounced for the higher frequency case.
}}
\label{fig:lc-vlow}
\end{figure}

\begin{figure}
\begin{center}
\hspace*{-0.1in}\includegraphics[scale=0.5]{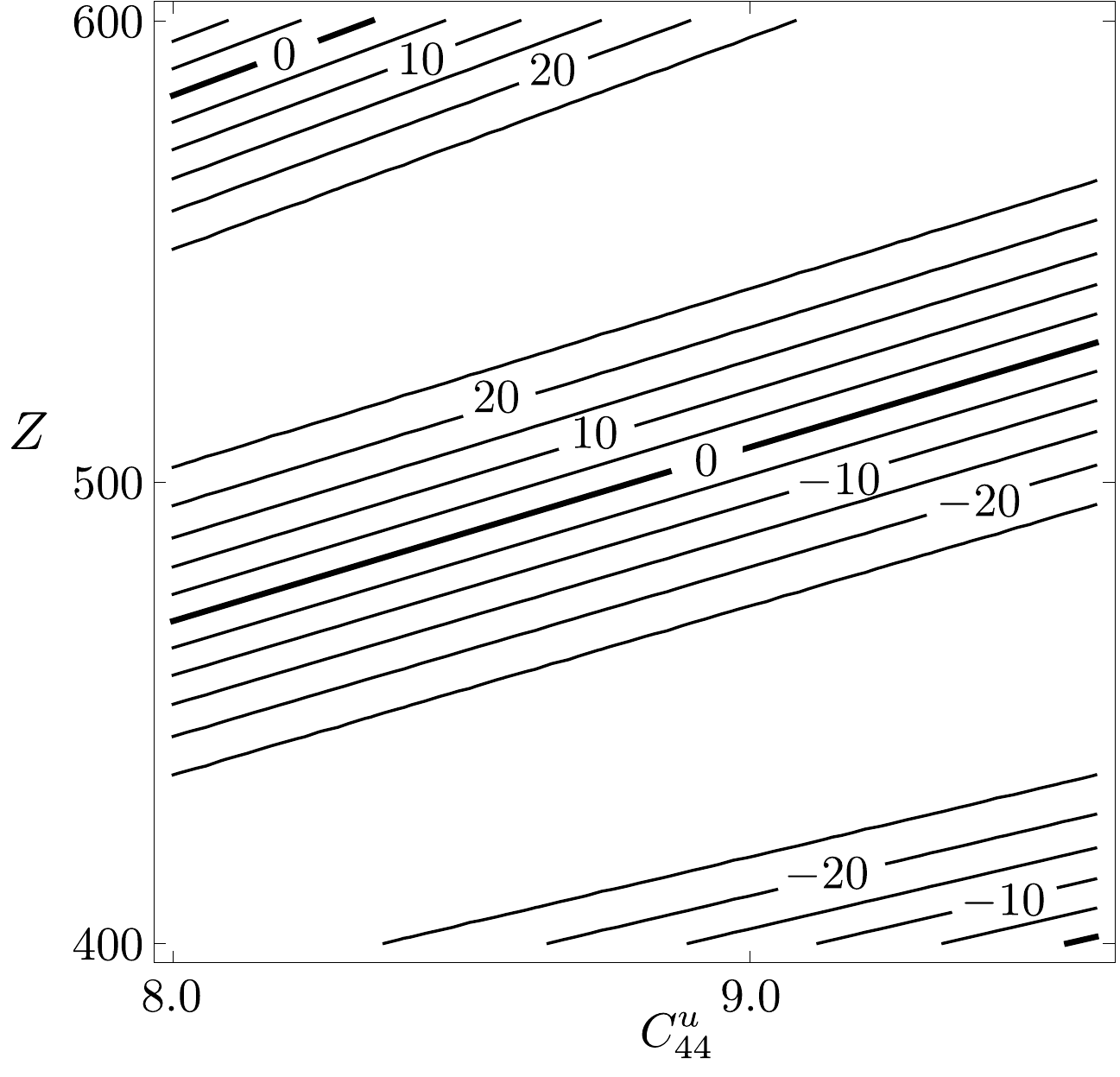}\hspace*{0.5in}\includegraphics[scale=0.5]{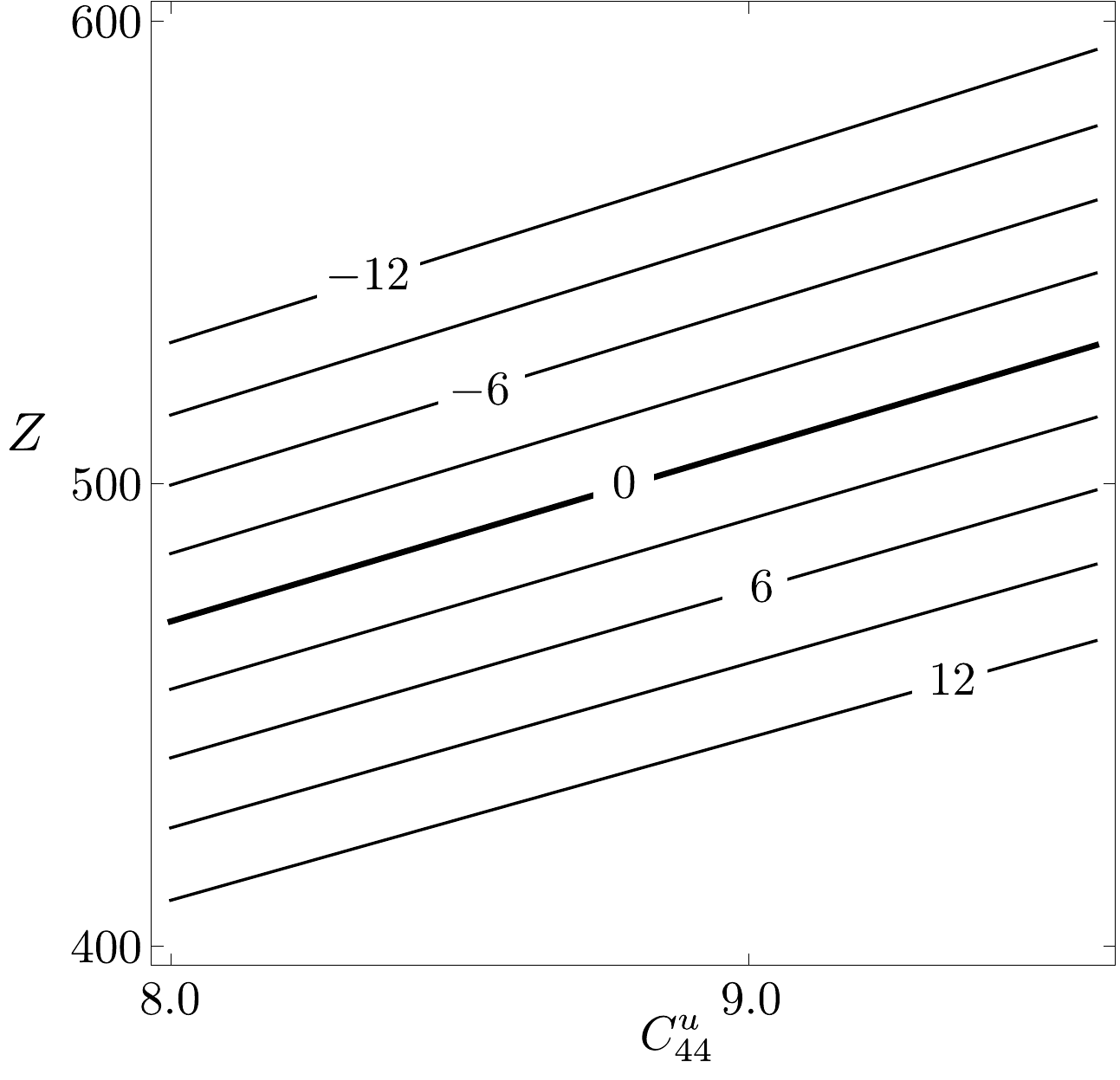}
%from love3x3contour-3.nb
\end{center}
\caption{\small{
$\mathrm{D}_\ell/10^{9}$\,, defined in expression~(\ref{eq:lovedisp2}), as a function of  $C_{44}^u$ and~$Z$\,, 
for
$\omega=60~{\rm s}^{-1}$ and 
$v_\ell=3958.53~{\rm m}/{\rm s}$\,, on the left,
and for
$\omega=15~{\rm s}^{-1}$ and 
$v_\ell=3997.01~{\rm m}/{\rm s}$\,, on the right.
The values on the horizontal axis are to be multiplied by~$10^9$\,.
In each case, there is sensitivity to both~$C_{44}^u$ and~$Z$\,.
}}
\label{fig:lcZ-vhigh}
\end{figure}

\begin{figure}
\begin{center}
\hspace*{-0.1in}\includegraphics[scale=0.5]{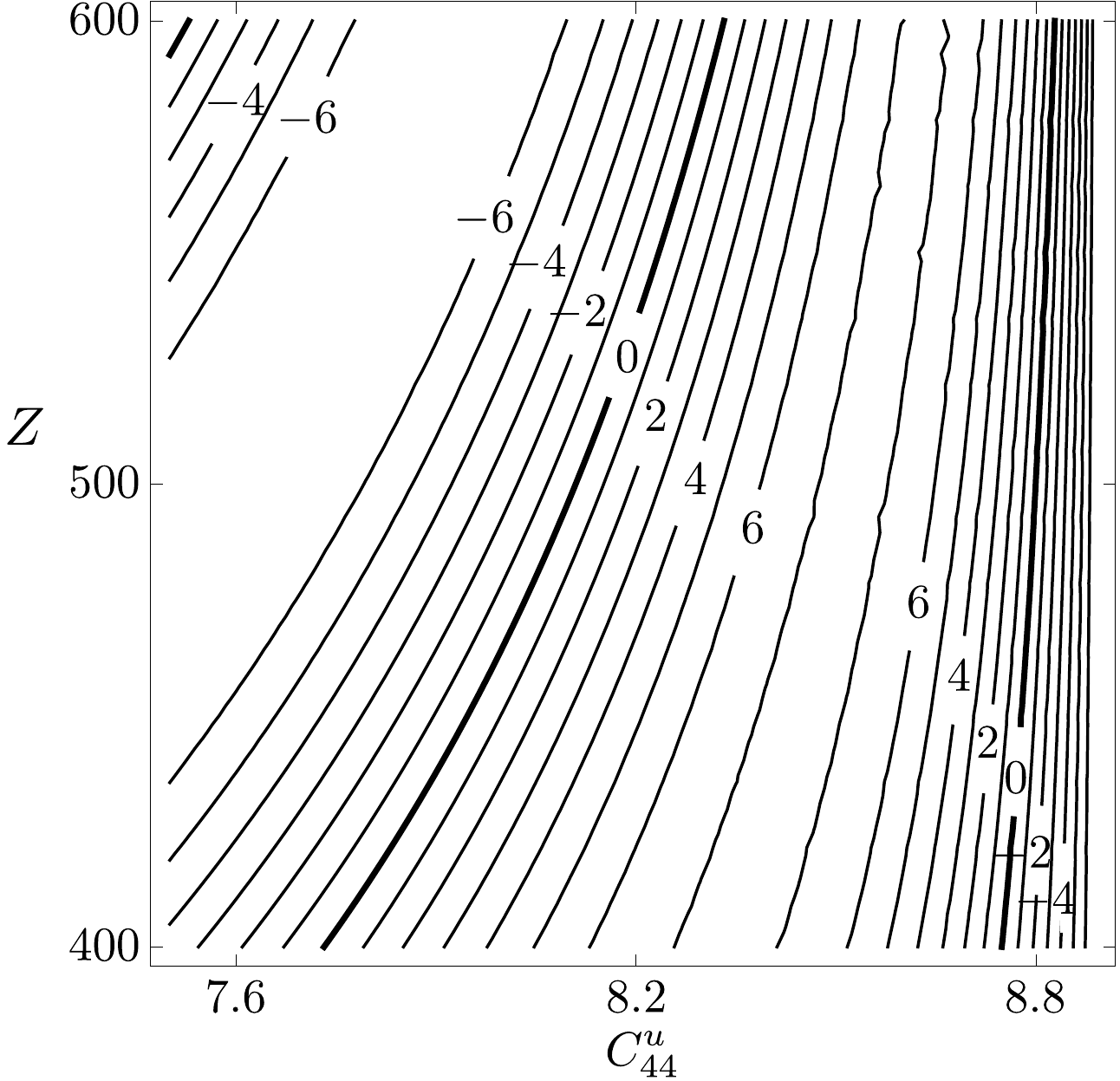}\hspace*{0.5in}\includegraphics[scale=0.5]{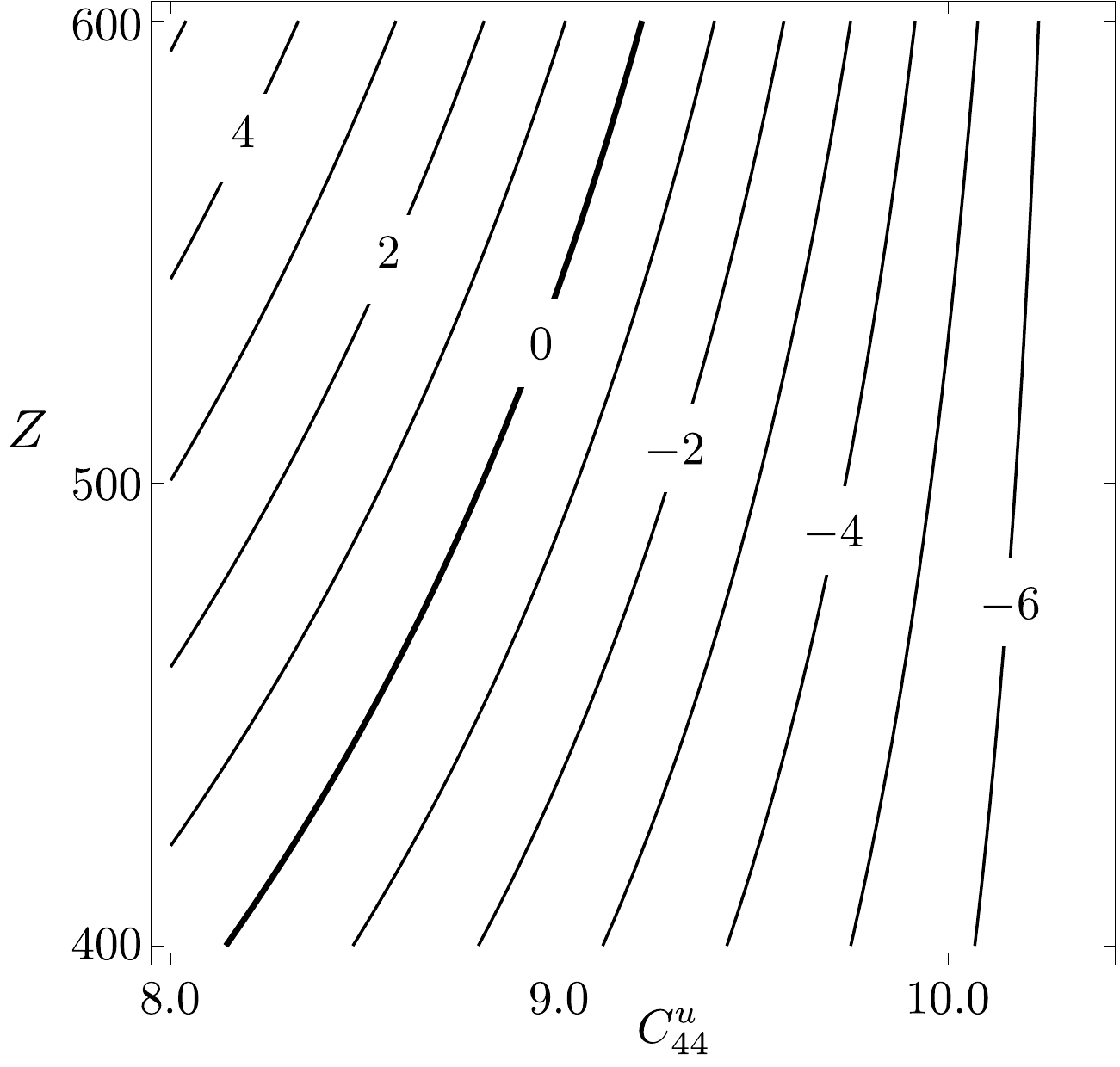}
%from love3x3contour-3.nb
\end{center}
\caption{\small{
$\mathrm{D}_\ell/10^{10}$\,, defined in expression~(\ref{eq:lovedisp2}), as a function of $C_{44}^u$ and $Z$\,, where
$\omega=60~{\rm s}^{-1}$ and
$v_\ell=2010.7~{\rm m}/{\rm s}$\,, and where
$\omega=15~{\rm s}^{-1}$ and
$v_\ell=2172.48~{\rm m}/{\rm s}$\,, for left and right, respectively.
In the left plot, the zero contour of interest is the near-vertical one on the right side of the plot, near the maximum~$C_{44}^u$\,.
The values on the horizontal axis are to be multiplied by~$10^9$\,.
In each case, there is sensitivity to both~$C_{44}^u$ and~$Z$\,, but the higher frequency case has less sensitivity to~$Z$\,.
}}
\label{fig:lcZ-vlow}
\end{figure}

In the contour maps of Figures~\ref{fig:lc-vhigh}--\ref{fig:lcZ-vlow}, the sensitivity of the dispersion relation is illustrated by the slope of the zero contour line, where a vertical line implies no sensitivity to the parameter on the vertical axis and a horizontal line implies no sensitivity to the parameter on the horizontal axis.
In other words, a parallel line does not restrict the values on the corresponding axis.
The zero contour line is the combination of elasticity parameters that result in a solution,  $D_\ell=0$.
Hence, from a vertical line we infer that the horizontal axis parameter is fixed but the vertical axis parameter is not constrained.
Our approach differs from the approach of Lucena and Taioli~\cite{Lucena}, who examine the response of the dispersion curves to shifts in parameter values.

Let us examine the numerical solutions of speed for the Love wave dispersion relation for the high-frequency and low-frequency cases.
We use these solutions to investigate the relative sensitivities of Love waves to the elasticity parameters in the upper layer and in the  lower halfspace, as well as to the elasticity parameter in the upper layer and to the layer thickness.

We begin by considering the layer and halfspace elasticity parameters.
For the high-speed case, we consider the fifth root of the left plot of Figure~\ref{fig:love}, which is~$v_\ell=3959~{\rm m}/{\rm s}$\,, and the second root of the right plot of Figure~\ref{fig:love}, which is~$v_\ell=3997~{\rm m}/{\rm s}$\,.
The left and right plots of Figure~\ref{fig:lc-vhigh} are the corresponding contour plots of~$D_\ell/10^{9}$ with varying~$C_{44}^u$ and~$C_{44}^d$\,. 
In both cases,~$D_\ell$ is sensitive to variations in both~$C_{44}^u$ and~$C_{44}^d$\,.
However, the right plot of Figure~\ref{fig:lc-vhigh} indicates a greater sensitivity to~$C_{44}^d$ for the lower frequency, which is tantamount to longer wavelength.

For the low-speed case, we observe different sensitivities.
Let us examine the first root of the left plot of Figure~\ref{fig:love}, which is~$v_\ell=2005~{\rm m}/{\rm s}$\,, and the first root of the right plot of Figure~\ref{fig:love}, which is~$v_\ell=2172~{\rm m}/{\rm s}$\,.
Following the corresponding plots in Figure~\ref{fig:lc-vlow}, we see that there are near vertical lines at~$C_{44}^u=0.88\times10^{10}$\,.
This indicates a sensitivity toward~$C_{44}^u$ but not toward~$C_{44}^d$\,, and we observe that it is more pronounced for the left plot.
Thus, for a given wavelength, a solution whose speed is closer to~$\beta^u$ is less sensitive to~$C_{44}^d$ than a solution with greater speed.

Next, we consider the layer elasticity parameter and layer thickness.
Using the high-speed roots of expression~(\ref{eq:lovedisp2}), we observe sensitivity to both~$C_{44}^u$ and~$Z$\,, in the high-frequency and low-frequency cases, which are depicted in Figure~\ref{fig:lcZ-vhigh}.
Using low-speed roots in Figure~\ref{fig:lcZ-vlow}, we see that there is less sensitivity to~$Z$ for higher frequencies than lower frequencies.
Note that, in the left plot of Figure~\ref{fig:lcZ-vhigh}, the groups of contours in the upper left and lower right are
due to the periodicity of~$D_\ell$ in~$Z$\,, as discussed is section~\ref{sec:superposition}, below.

Additionally, the two plots of Figure~\ref{fig:lcZ-vlow}, along with the left plot of Figure~\ref{fig:lc-vlow}, approach a maximum value of~$C_{44}^u$\,, whereas the two plots of Figure~\ref{fig:lcZ-vhigh} do not.
To understand this behaviour, we recall that, as a consequence of the allowable range of Love-wave speeds, $\beta^u<v_\ell<\beta^d$\,, there is a maximum value of~$C_{44}^u$ whose value is~$v_\ell^2\rho^u$\,.
%%%%%%%%%%%%%%%%%%%%%%%%%%%%%%%%%%%%%%%%%%%
\subsection{Love wave as superposition of $SH$ waves}
\label{sec:superposition}
%%%%%%%%%%%%%%%%%%%%%%%%%%%%%%%%%%%%%%%%%%%
The ridge and valley behaviour, shown in Figures~\ref{fig:lcZ-vhigh} and~\ref{fig:lcZ-vlow}, can be explained by examining the nonzero component of the displacement,
\begin{equation*}
u_2^u(x_1,x_3,t) =
C_1\exp{\left(-\iota\kappa s^u x_3\right)} \exp[\iota(\kappa x_1-\omega t)] + C_2\exp{\left(\iota\kappa s^u x_3\right)}\exp[\iota(\kappa x_1-\omega t)]\,,
\end{equation*}
where $s^u:=\sqrt{(v_\ell/\beta^u)^2-1}$\,.
This expression can be interpreted as a superposition of two~$SH$ waves within the elastic layer.
Both waves travel obliquely with respect to the surface and interface; one wave travels upwards, the other downwards.
Their wave vectors are~$\overline{k}_{\pm}:=(\kappa, 0, \pm \kappa s^u)$\,.
Thus, since 
\begin{equation*}
|\overline{k}_\pm|=\sqrt{\kappa^2+(\kappa\,s^u)^2}\,,
\end{equation*}
we have 
\begin{equation*}
|\overline{k}_\pm|=\kappa\sqrt{1+(s^u)^2}=\kappa\,\frac{v_\ell}{\beta^u}\,,
\end{equation*}
where  
\begin{equation*}
\frac{\beta^u}{v_\ell}=\frac{\kappa}{|k_\pm|}=\sin \theta\,,
\end{equation*}
with~$\theta$ representing for the angle between~$\overline{k}_{\pm}$ and the~$x_3$-axis.
Thus,~$\theta$ is the angle between the~$x_3$-axis and a wavefront normal, which implies that it is the propagation direction of a wavefront.
Hence, for the case where~$v_\ell$ is only slightly larger than~$\beta^u$\,, it follows that both the upgoing and downgoing~$SH$ waves propagate nearly horizontally.
In other words, their propagation directions are nearly parallel to the interface.
In such a case, the resulting Love wave is less sensitive to the material properties below the interface than for the case of~$\beta^u/v_\ell\ll 1$\,.

Depending on the propagation direction,~$\theta$\,, of the~$SH$ waves, there is a distinction in the sensitivity to the material properties of the halfspace.
This distinction is more pronounced for short wavelengths, in comparison with the layer thickness.
For any angle, the longer the wavelength the more sensitivity of the wave to material properties below the interface, and thus the distinction---as a function of the propagation direction---is diminished.%

Also, the superposition helps us understand the requirement of~$\beta^u<v_\ell<\beta^d$\,.
The lower limit, which we can write as~$\sin \theta=\beta^u/\beta^u=1$\,, corresponds to waves that propagate along the~$x_1$-axis, and hence, do not exhibit any interference associated with the presence of the horizontal surface or interface.
Formally, the lower limit is required by the real value of the sine function.

The upper limit is introduced to ensure an exponential amplitude decay in the halfspace.
Herein, we can write the upper limit as~$\sin \theta=\beta^u/\beta^d$\,, which---in general---implies that the~$SH$ waves propagate obliquely. 
In the extreme case, if~$\beta^d\to\infty$\,, the waves propagate vertically.
Also, this case is tantamount to total internal reflection, since it corresponds to a rigid halfspace,~$C_{44}^d\to\infty$\,; such a case is discussed by Slawinski~\cite[Section~6.3.1]{SlawinskiWS2016}.
%%%%%%%%%%%%%%%%%%%%%%%%%%%%
\subsection{Optimum frequency for layer-thickness determination}
\label{sec:optimum}
%%%%%%%%%%%%%%%%%%%%%%%%%%%%
In this section, we search for the best frequency for inferring the depth of the interface.
Given the values of the elasticity parameters and layer thickness, for a specific frequency, we deduce the corresponding propagation speed.
For a given mode, this speed depends on~$Z$\,, which means that~$Z$ is a function of speed. 
Since, experimentally, the speed is measured with only finite accuracy, we inherit that finite accuracy for any inference of the value of~$Z$\,. 

As a measure of  sensitivity of~$Z$ to errors in speed, we take the derivative,~${\rm d}Z/{\rm d}v$\,, or, more precisely, its ratio to~$Z$\,, namely,
\begin{equation}
\tilde{A}:=\frac{1}{Z}\frac{{\rm d}Z}{{\rm d}v}\,.
\label{eq:Zvsensitivity}
\end{equation}
For convenience, we write relation (\ref{eq:lovedisp2}) using dimensionless quantities.
If we let
\begin{equation}
\alpha:=\sqrt{\frac{C_{44}^d\rho^u}{C_{44}^u\rho^d}}=\frac{\beta^d}{\beta^u}>1\,,\quad
E:=\sqrt{\frac{C_{44}^d\rho^d}{C_{44}^u\rho^u}}\,,\quad
\theta:=\frac{v_\ell}{\beta^u}\,,\quad
\nu:=\frac{\omega Z}{\beta^u}\,,
\label{eq:dimvar}
\end{equation}
then  relation (\ref{eq:lovedisp2}) becomes
\begin{equation}
F(\nu,\theta):=\sqrt{\theta^2-1}\sin\left(\nu\sqrt{1-\frac{1}{\theta^2}}\right)
-E\sqrt{\alpha^2-\theta^2}\cos\left(\nu\sqrt{1-\frac{1}{\theta^2}}\right)=0\,,
\label{eq:F}
\end{equation}
where~$1<\theta<\alpha$ and~$\nu>0$\,.
Using the Implicit Function Theorem, the following properties are shown.
\begin{itemize}
\item[$\circ$]
$F(\nu,\theta)=0$ defines locally~$\theta$ as a function of~$\nu$\,, and, reciprocally,~$\nu$  as a function of~$\theta$\,.
\item[$\circ$]
$F(\nu,\theta)=0$ has a solution~$1< \theta< \alpha$ for any~$\nu>0$\,.
The smallest such solution is the fundamental mode; it is a smooth and strictly decreasing function of~$\nu$; it is invertible and defines~$\nu$ as a function of~$\theta$\,.
\item[$\circ$]
Higher modes start at~$\nu_k>0\,,\,k=1,2,\dots,$ and define~$\theta$ as strictly decreasing functions of~$\nu$\,.
\end{itemize}
The Implicit Function Theorem is also used in a related study by Novotn\'y~\cite{Novotny} except his dispersion relation remains in its tangent form.

Following the definition of~$ \tilde{A}$\,, in expression~(\ref{eq:Zvsensitivity}), it is convenient to measure speed in units of~$\beta^u$ so that~$\theta$ is used instead of~$v$\,. 
Thus, we define the dimensionless coefficient as
\begin{equation} 
A:=\frac{1}{Z}\frac{ {\rm d} Z}{{\rm d} \theta}\,.
\label{eq:A}
\end{equation}
$Z(\theta)$\,, which appears in definition~(\ref{eq:A}), is given by
\begin{equation}
Z(\theta):=\frac{\beta^u}{\omega}\,\nu(\theta)\,,
\end{equation}
and~$\nu(\theta)$ is a mode defined by expression~(\ref{eq:F}). 
Hence,
\begin{equation}
A=\frac{1}{\nu}\frac{{\rm d}\nu}{{\rm d}  \theta}= \frac{1}{\nu}\frac{1}{\frac{{\rm d} \theta}{{\rm d} \nu}}\,,\qquad\mbox{where}\qquad
\frac{{\rm d} \theta}{{\rm d}  \nu}=-\frac{\frac{\partial F}{\partial \nu}}{\frac{\partial F}{\partial \theta}}\,.
\end{equation}
To calculate and plot~$A(\omega)$\,, given~$Z$\,, we calculate~$\nu=(\omega Z)/\beta^u$\,,  given~$Z$ and~$\omega$\,, and numerically find the minimum solution,~$\theta(\nu)$\,, for~$F=0$\,.

Calculating the partial derivatives and using
\begin{equation}
\sin\left(\nu\sqrt{1-\frac{1}{\theta^2}}\right)=
\frac{E\sqrt{\alpha^2-\theta^2}\cos\left(\nu\sqrt{1-\frac{1}{\theta^2}}\right)}{\sqrt{\theta^2-1}}\,,
\label{eq:sin}
\end{equation}
from expression~(\ref{eq:F}), we obtain
\begin{equation}
A =
-\frac{1}{\nu}\frac{\left(\alpha^2-1\right) E \theta^3+E^2 \nu (\alpha^2-\theta^2)^{3/2}+\nu \left(\theta^2-1\right) \sqrt{\alpha^2-\theta^2}}{\theta \left(\theta^2-1\right) 
\sqrt{\alpha^2-\theta^2}\left(\theta^2-1 + E^2( \alpha^2-\theta^2)\right)}\,,
\label{eq:rel}
\end{equation}
which is the relative dimensionless sensitivity coefficient.

\begin{figure}
\begin{center}
\hspace*{-0.1in}\includegraphics[scale=0.5]{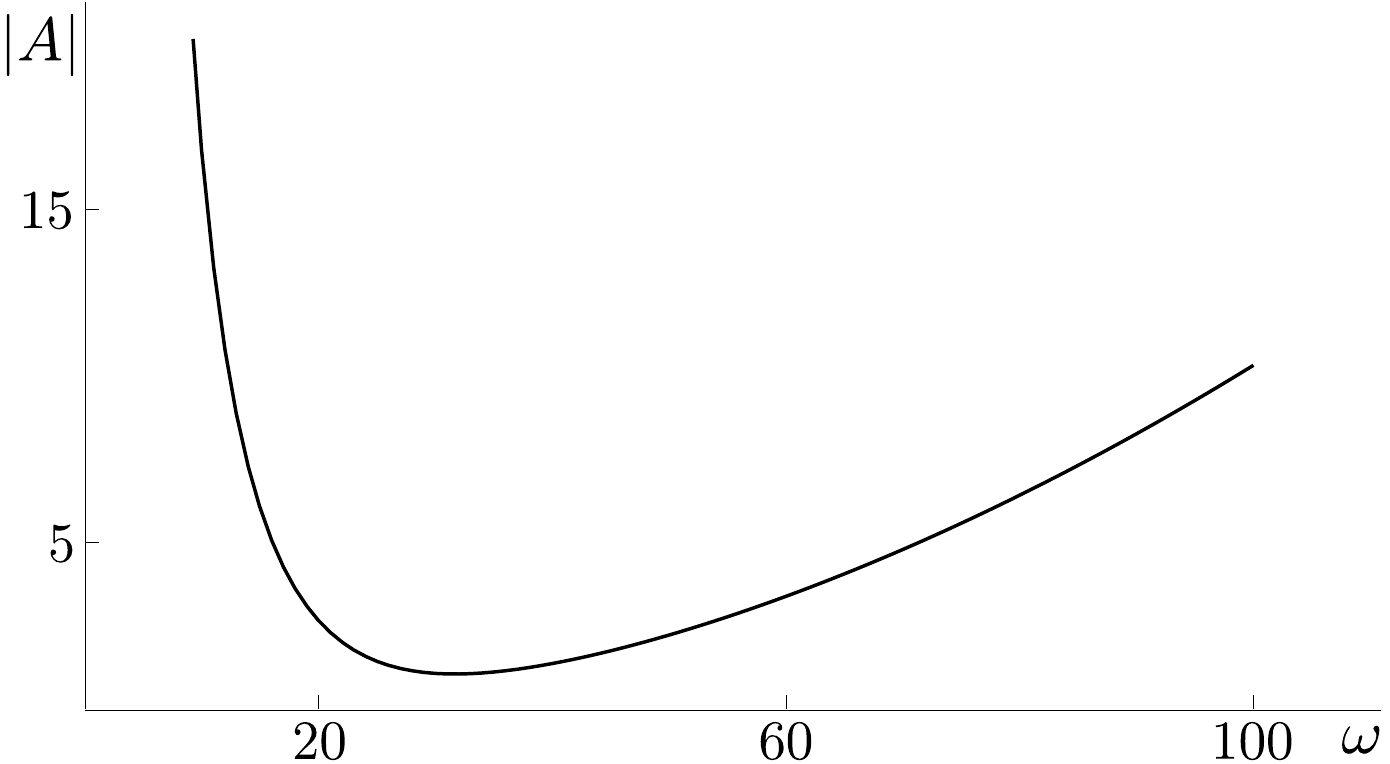}\hspace*{0.5in}\includegraphics[scale=0.5]{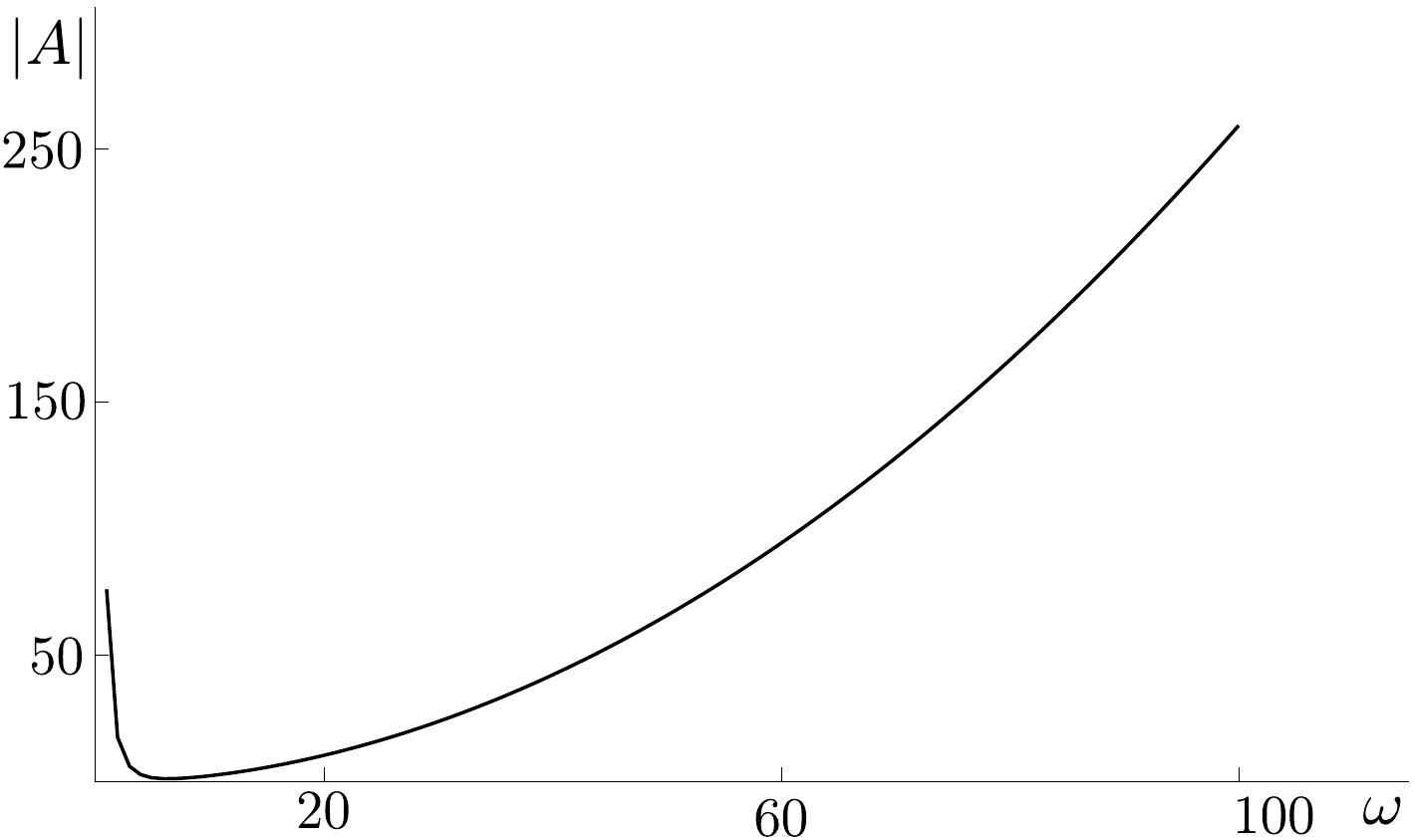}
%from sensitivity2.nb
\end{center}
\caption{\small{
The absolute value of the relative dimensionless sensitivity coefficient,~$|A|$\,, is plotted against~$\omega$\,, for
$Z=100~{\rm m}$ on the left, and for~$Z=500~{\rm m}$ on the right.
Both plots begin at~$\omega=1\,{\rm s}^{-1}$ due to the asymptote at~$\omega=0$\,.
Due to the different vertical scales, the minimum~$|A|$ on each plot is the same.
}}
\label{fig:sensitivity2}
\end{figure}

In Figure~\ref{fig:sensitivity2}, we examine the relationship between~$|A|$ and~$\omega$ for different layer thicknesses.
In the left plot, which is the case of~$Z=100~{\rm m}$\,, we see that there is a minimum $|A|$ at about~$\omega_0=30~{\rm s}^{-1}$\,.
In the right plot, which is the case for~$Z=500~{\rm m}$\,, the same minimum $|A|$ occurs at about~$\omega_0=7~{\rm s}^{-1}$\,.

We wish to calculate the minimal value of~$|A|$  to find the frequency,~$\omega_0$\,, where the layer thickness,~$Z$\,, is least sensitive to variations in speed.
To find that value of~$|A|$\,, we solve for~${\rm d} A/{\rm d} \nu=0$ or, equivalently, 
\begin{equation}
\frac{{\rm d}}{{\rm d} \nu}\left(\frac{1}{A}\right)=
\frac{{\rm d}}{{\rm d} \nu}\left(\nu \frac{{\rm d} \theta}{{\rm d} \nu}\right)=\frac{{\rm d} \theta}{{\rm d} \nu}+ \nu \frac{{\rm d}^2 \theta}{{\rm d} \nu^2}=\theta'+\nu\theta''=0\,.
\label{eq:thpp}
\end{equation}
By differentiating $F(\nu,\theta(\nu))=0$ twice  with respect to $\nu$\,, we get
\begin{equation*}
\frac{\partial^2 F}{\partial \nu^2}+ 2 \frac{\partial^2 F}{\partial \theta \partial \nu}\theta'+ \frac{\partial^2 F}{\partial \theta^2}(\theta')^2 + 
\frac{\partial F}{\partial \theta }\theta''=0\,,
\end{equation*}
where prime denotes derivative with respect to~$\nu$\,.
In this manner, $\theta'+ \nu \theta''=0$ in expression~(\ref{eq:thpp}) can be written in terms of~$F$ as
\begin{equation}\label{eq:F2}
F_2(\nu,\theta)=- 2 \frac{\partial^2 F}{\partial\theta \partial\nu}\frac{\partial F}{\partial\theta}+ 
\frac{1}{\nu}\left( \frac{\partial F}{\partial \theta}\right)^2 +
\frac{\partial^2 F}{\partial\theta^2 } \frac{\partial F}{\partial\nu}=0\,,
\end{equation}
where we use
\begin{equation*}
\frac{\partial^2 F}{\partial\nu^2}=\frac{1-\theta^2}{\theta^2} F=0\,.
\end{equation*}
\begin{figure}
\begin{center}
\hspace*{-0.1in}\includegraphics[scale=0.5]{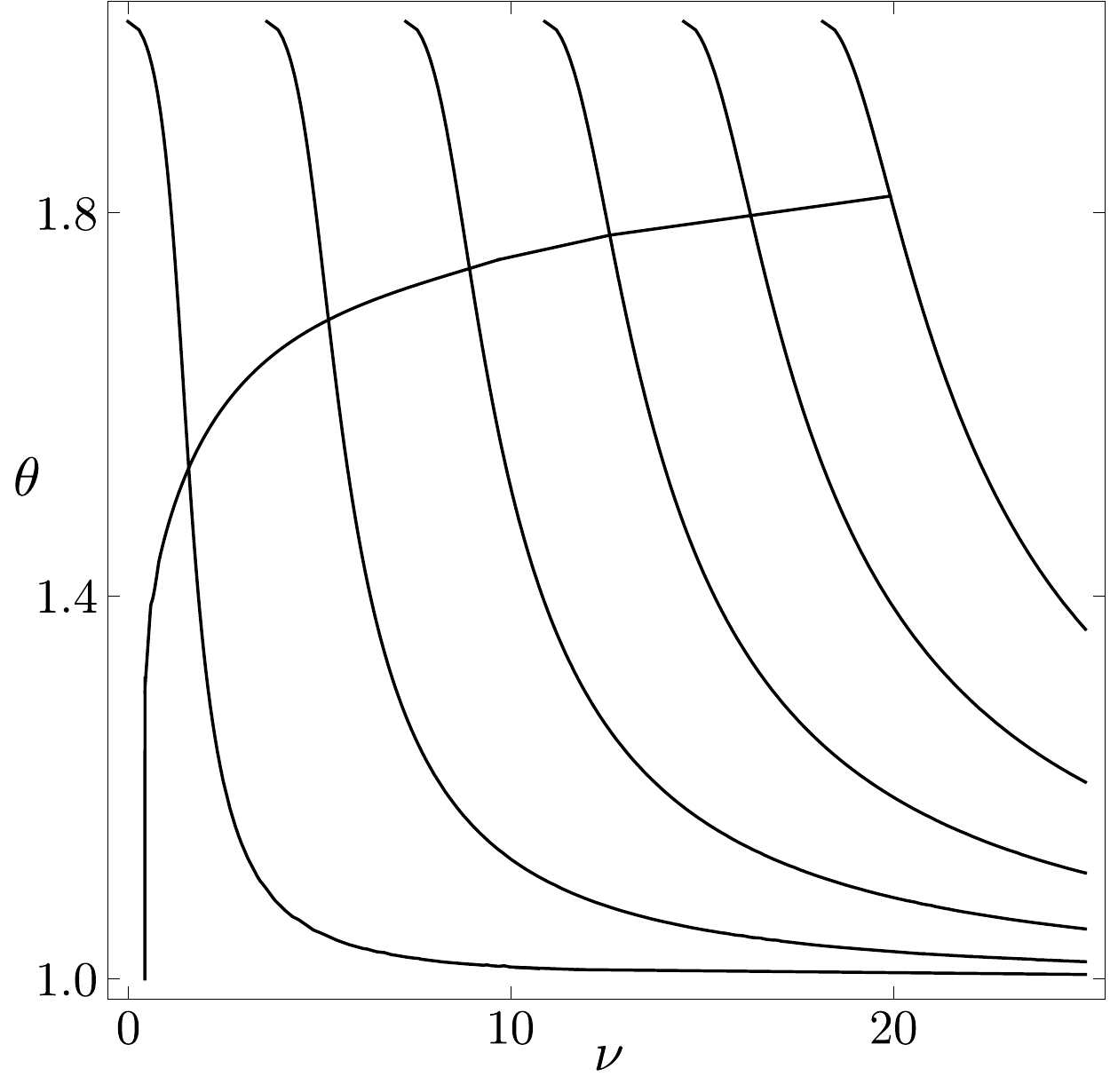}\hspace*{0.5in}\includegraphics[scale=0.5]{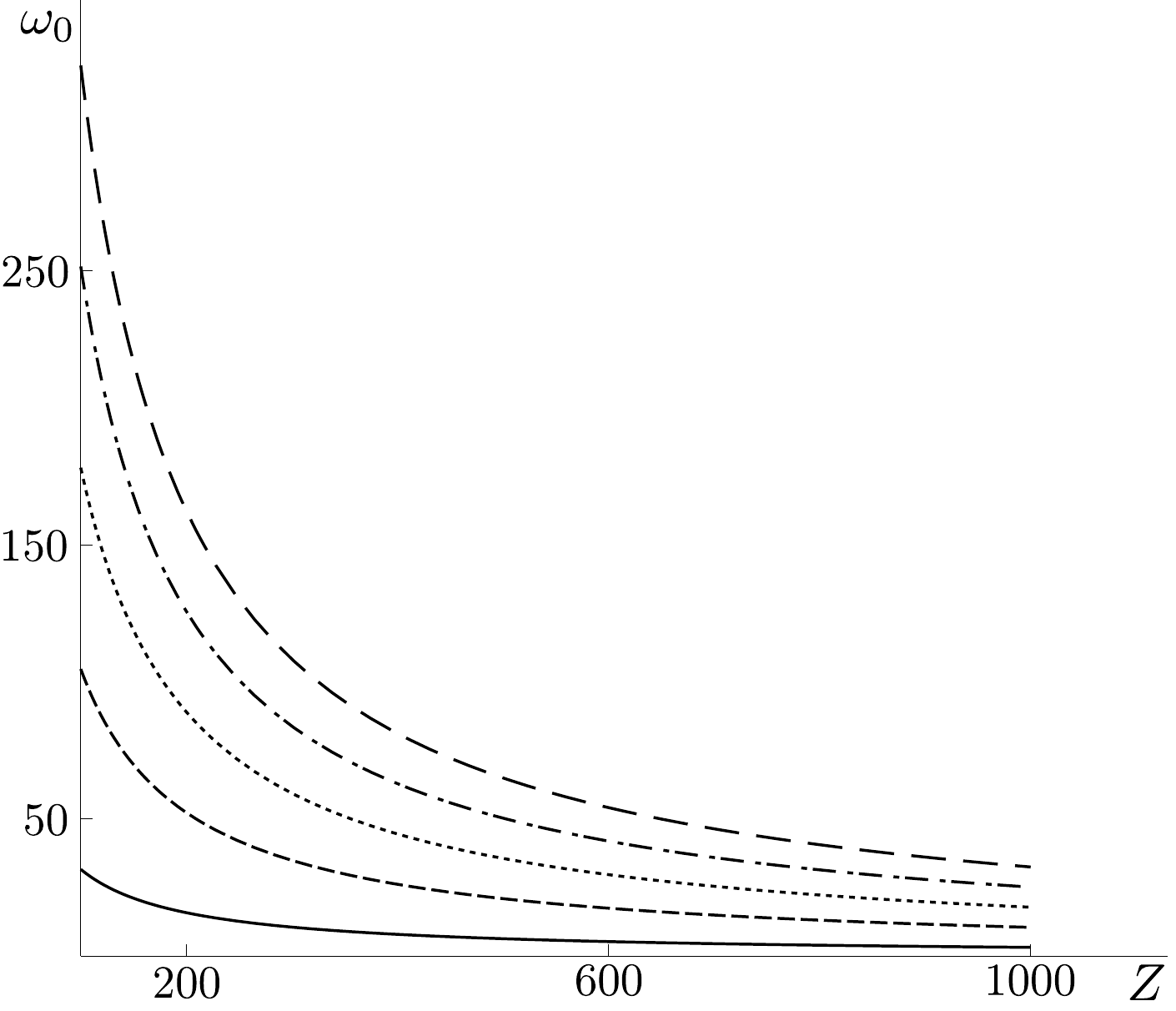}
%from sensitivity3.nb
\end{center}
\caption{\small{
The left plot depicts intersecting zero contours of~$F(\nu,\theta)$ and~$F_2(\nu,\theta)$\,, where the curve ascending from~$(0,1)$ is the zero contour of~$F_2$\,.
The right plot depicts the optimum frequency~$\omega_0$ plotted versus~$Z$\,.
}}
\label{fig:sensitivity3}
\end{figure}
Evaluating the partial derivatives and using expression~(\ref{eq:sin}), we obtain
{%\small
\begin{eqnarray}
F_2(\nu,\theta)&=&\frac{\cos ^2\left(\frac{\nu \sqrt{\theta^2-1}}{\theta}\right) 
}{\theta^5 \left(\theta^2-1\right)^2(\alpha^2-\theta^2)}
\left\{\frac{\theta \left(\left(\alpha^2-1\right) E \theta^3+E^2 \nu (\alpha^2-\theta^2)^{3/2}+\nu \left(\theta^2-1\right) \sqrt{\alpha^2-\theta^2}\right)^2}{\nu}\right.\nonumber\\
&+&
\frac{1}
{\sqrt{\alpha^2-\theta^2}}
\left[
\left(E^2 (\alpha^2-\theta^2)+\theta^2-1\right) \left(\left(\theta^2-1\right) \left(\alpha^4 E \nu^2-\alpha^2 \theta \left(\nu \left(\theta^2-2\right) \sqrt{\alpha^2-\theta^2}
\right.\right.\right.\right.\nonumber\\
& &\quad+\left.\left.
E \left(2 \nu^2 \theta-\theta^5+\theta^3\right)\right)
+\nu \theta^3 \left(\left(\theta^2-2\right) \sqrt{\alpha^2-\theta^2}+E \nu \theta\right)\right)
-E (\alpha^2-\theta^2)^{3/2} 
\nonumber\\
& &\qquad\times\left.\left.
\left(\alpha^2 E \nu \left(3 \theta^2-2\right) \theta+\sqrt{\alpha^2-\theta^2} \left(\nu^2 \left(\theta^2-1\right)+\theta^4\right)-E \nu \theta^5\right)\right)\right]\nonumber\\
&-&
\left[
2 \theta \left(E^2 \left(\alpha^2-\theta^4\right)+\theta^4-1\right) \left(\alpha^4 E^2 \nu+\alpha^2 \left(E \theta^3 \sqrt{\alpha^2-\theta^2}\right.\right.\right.
\label{eq:F2long}\\
& &\quad+\left.\left.\nu \left(-2 E^2 \theta^2+\theta^2-1\right)\Big)+\theta^2 \left(-E \theta \sqrt{\alpha^2-\theta^2}+\left(E^2-1\right) \nu \theta^2+\nu\right)\right)\right]\Bigg\}=0\,.
\nonumber
\end{eqnarray}}
Using the system of equations formed by expressions~(\ref{eq:F}) and~(\ref{eq:F2long}), we obtain~$\nu_0$ for which the absolute value of~$A$ is minimal for a given mode.

The intersection points of~$F(\nu,\theta)$ and~$F_2(\nu,\theta)$\,, on the left plot of Figure~\ref{fig:sensitivity3}, correspond to solutions,~$(\nu_0,\theta_0)$\,, for the first seven modes of~$A$ in expression~(\ref{eq:rel}).   
In view of the intrinsic nonlinearity, we estimate the coordinates of the intersection points and use them as starting points in a numerical search.
The values of $(\nu_0,\theta_0)$\,, for the first seven modes, are
$(1.58172,1.53806)$\,,
$(5.23429,1.68935)$\,,
$(8.90672,1.74515)$\,,
$(12.5772,1.7779)$\,,
$(16.2443,1.80044)$\,,
$(19.9085,1.81731)$ and
$(23.57,1.83059)$\,.
Using the values in equation~(\ref{eq:rel}), we obtain values of~$|A|_{\rm min}$ for the first seven modes.
They are 
$1.05$\,,
$0.555$\,,
$0.445$\,,
$0.392$\,,
$0.360$\,,
$0.338$ and
$0.321$\,,
and are constant for all~$Z$ for a given mode.
This is shown in Figure~\ref{fig:sensitivity2} for~$|A|_{\rm min} = 1.05$
for both~$Z=100$ and~$Z=500$\,.

On the right plot of Figure~\ref{fig:sensitivity3}, the relationship between the optimum frequency,~$\omega_0$\,, and~$Z$ for the first five modes, is illustrated, where the solid black line closest to the horizontal axis corresponds to the first mode.
Using the definition of~$\nu$ from expression~(\ref{eq:dimvar}), we are able to compare values of~$\omega_0$ with those estimated in Figure~\ref{fig:sensitivity2}.
Using the fundamental mode,~$\nu_0=1.58172$\,, we calculate, for $Z=500$ a value of~$\omega_0=6.33$\,, which compares favourably to $\omega_0=7$\,, which is estimated from the right plot of Figure~\ref{fig:sensitivity2}.
For~$Z=100$\,, we obtain a value of~$\omega_0=31.6$\,, which compares
favourably to the~$\omega_0=30$  estimated from the left plot of Figure~\ref{fig:sensitivity2}.

The study presented in this section, which gives the optimum frequency for measuring~$Z$ for a given mode for Love waves, could be repeated for other quantities, such as the elasticity parameters, holding~$Z$ fixed.
A similar study might be performed for quasi-Rayleigh waves.
%%%%%%%%%%%%%%%%%%%%%%%%%%%%
\section{Quasi-Rayleigh waves}
\label{sec:quasirayl}
%%%%%%%%%%%%%%%%%%%%%%%%%%%%
\subsection{Material properties}
%%%%%%%%%%%%%%%%%%%%%%%%%%%%
To consider the quasi-Rayleigh wave within the model of material properties discussed in Section~\ref{sec:LovWav}, we also need to specify~$C_{11}^u$ and~$C_{11}^d$\,, which are the elasticity parameters  for the layer and the halfspace that do not appear in the Love-wave equations.
Hence, the corresponding~$P$-wave propagation speeds are~$\alpha^u=\sqrt{C_{11}^u\strut/\rho^u}$ and~$\alpha^d=\sqrt{\strut C_{11}^d/\rho^d}$\,.

We invoke the wave number~$\kappa=\omega/v_{r}$\,, where~$v_{r}$ is the propagation speed.
Herein, this speed corresponds to the quasi-Rayleigh wave.
For a notational convenience, we let
\begin{equation}
r^u:=\sqrt{\frac{v_{r}^2}{(\alpha^u)^2}-1}\,,\quad
s^u:=\sqrt{\frac{v_{r}^2}{(\beta^u)^2}-1}\,,\quad
r^d:=\sqrt{1-\frac{v_{r}^2}{(\alpha^d)^2}}\,,\quad
s^d:=\sqrt{1-\frac{v_{r}^2}{(\beta^d)^2}}\,.
\label{eq:rusurdsd}
\end{equation}

Following a laborious process, shown in~\ref{app:qRwaves} and~\ref{app:disprel}, we obtain the dispersion relation, which is expressed as the determinant of the coefficient matrix,
\begin{equation}
\label{eq:ourdet}
\mathrm{D}_r := \det[M_r] =
4C^u_{44} \det\left[
\begin{array}{cc} 
s^u X & s^u S\\r^u T & r^u Y
\end{array}
\right]\,,
\end{equation}
where~$X,Y,S,T$ are
{%\small
\begin{equation*}
X :=
\left[(s^u)^2-1\right]\left[-(v_{r}^2 q+ 2 p) B' +2 p r^d \cos b'\right] +
2 \left[r^u ( 2 p -v_{r}^2 \rho^d)\sin a' + r^d( 2 p + v_{r}^2 \rho^u)\cos a'\right]\,,
\end{equation*}
\begin{equation*}
Y := \left[(s^u)^2-1\right]\left[(v_{r}^2 q+ 2 p) A' -2 p s^d \cos a'\right] +
2 \left[- s^d ( 2 p +v_{r}^2 \rho^u)\cos b' - s^u( 2 p - v_{r}^2 \rho^d)\sin b'\right],
\end{equation*}
\begin{equation*}
S := \left[(s^u)^2-1\right]\left[-(v_{r}^2\rho^u+ 2 p)s^d B' + (2 p-v_{r}^2 \rho^d) \cos b'\right] +
2 \left[ ( 2 p +v_{r}^2 q)\cos a' + 2 p r^u s^d \sin a'\right],
\end{equation*}
\begin{equation*}
T := \left[(s^u)^2-1\right]\left[r^d (v_{r}^2 \rho^u + 2 p) A' -(2 p-v_{r}^2 \rho^d) \cos a'\right] - 
2 \left[( 2 p +v_{r}^2 q)\cos b' +2  s^u r^d p \sin b'\right],
\end{equation*}}
with~$q, p, a', b', A', B'$ given by
\begin{equation*}
q:=\rho^u-\rho^d\,,\quad 
p:=C^d_{44}-C^u_{44}\,,\quad 
a':=\kappa r^u Z\,,\quad 
b':=\kappa s^u Z\,,
\end{equation*}
\begin{equation*}
A':=
\begin{dcases}
\frac{\sin a'}{r^u}&r^u\neq0\\
\kappa Z&r^u=0
\end{dcases}\,,
\qquad
B':=
\begin{dcases}
\frac{\sin b'}{s^u}&s^u\neq0\\
\kappa Z&s^u=0
\end{dcases}\,,
\end{equation*}
where both~$A'$ and~$B'$ are real, regardless of whether or not~$r^u$ and~$s^u$ are real or imaginary.
In accordance with l'H\^opital's rule, both~$A'$ and~$B'$ are equal to~$\kappa Z$\,, for~$r^u=0$ and~$s^u=0$\,, in the limit sense.
From our calculations, we find that that~$X,Y,S,T$ are real,~$\mathrm{D}_r=0$ for~$s^u=0$ or~$r^u=0$\,, and  whether~$\mathrm{D}_r$ is real or imaginary depends only on whether the product~$s^u\,r^u$ is real or imaginary,
which depends on the value of~$v_{r}$\,.
We express the latter dependance as

\begin{equation}
{\rm D}_{r} :
\begin{cases}
\begin{tabular}{ll}
Real & for\quad $v_{r}\in(0,\beta^{u})\cup(\alpha^{u}\,, \beta^{d})$ \\
Imaginary & for\quad $v_{r}\in(\beta^{u}\,, \alpha^{u})$
\end{tabular}
\end{cases}
\,.
\label{eq:DrDomains}
\end{equation}
Also, there are body-wave solutions for~$r^u=0$ and for~$s^u=0$\,, which means that~$v_r=\alpha^u$\,, and~$v_r=\beta^u$\,, respectively.
However, from analyses of equations~(\ref{eq:QRModBC1})--(\ref{eq:QRModBC4}), (\ref{eq:QRModBC5}) and~(\ref{eq:QRModBC6}), we conclude that their displacements are zero, and hence, these solutions are trivial.

\begin{figure}
\begin{center}
\includegraphics[scale=0.5]{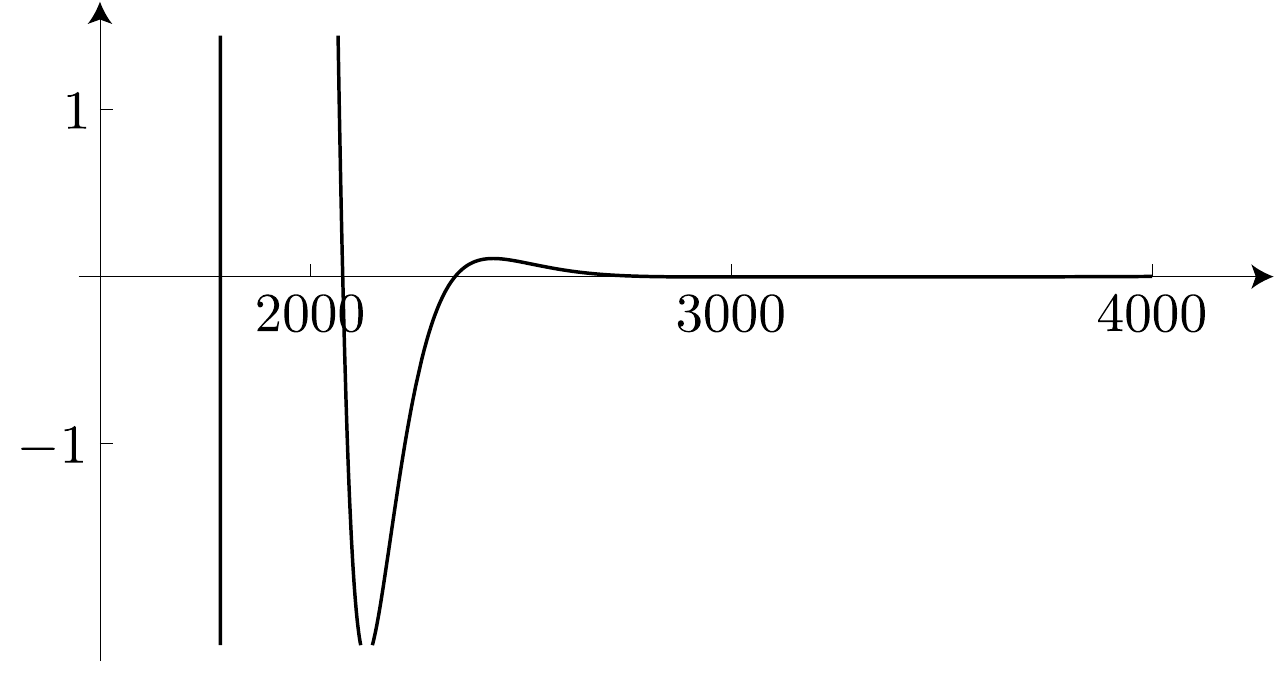}\hspace*{0.5in}\includegraphics[scale=0.5]{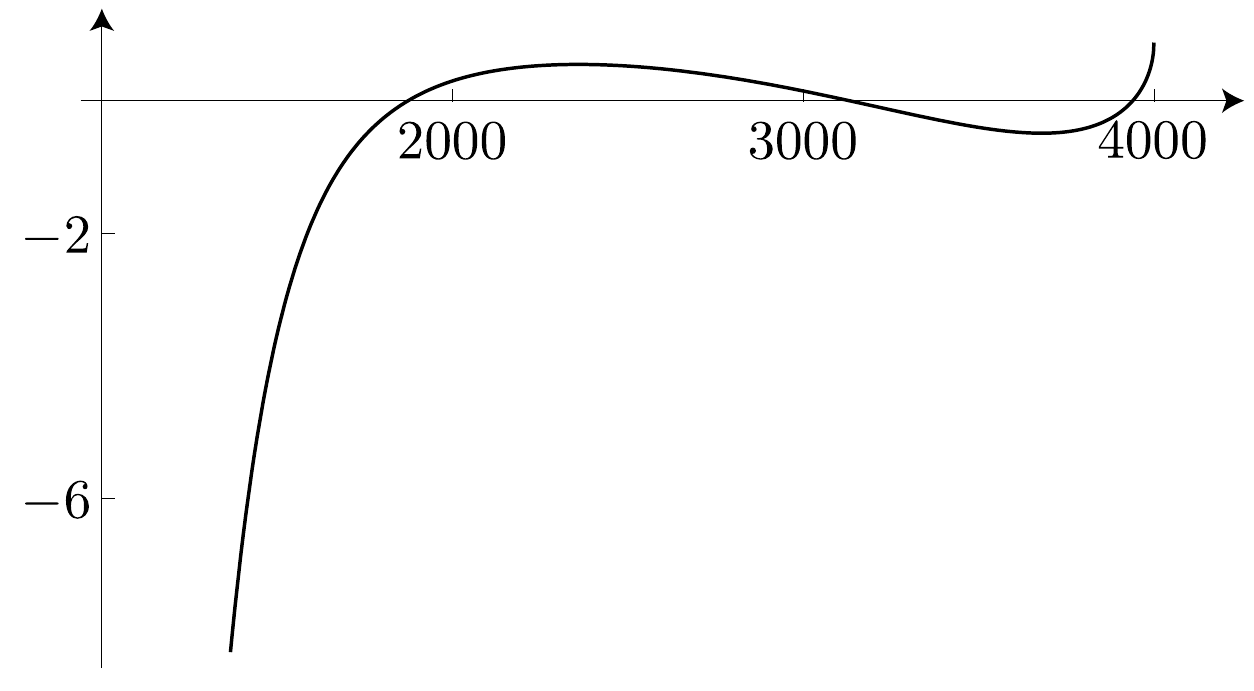}
\end{center}
\caption{\small{
$\mathrm{D}_{r2}$\,, defined in expression~(\ref{eq:Dr2}), as a function of speed~$v_{r}$\,. 
On the left, for~$\omega=60~{\rm s}^{-1}$\,, there are seven roots:~$v_{r}=1786~{\rm m}/{\rm s}$\,,  
$v_{r}=2077~{\rm m}/{\rm s}$\,,~$v_{r}=2343~{\rm m}/{\rm s}$\,,
$v_{r}=2869~{\rm m}/{\rm s}$\,,~$v_{r}=3075~{\rm m}/{\rm s}$\,,
$v_{r}=3288~{\rm m}/{\rm s}$ and
$v_{r}=3705~{\rm m}/{\rm s}$\,.
On the right, for~$\omega=15~{\rm s}^{-1}$\,, there are three roots:
$v_{r}=1869~{\rm m}/{\rm s}$\,,~$v_{r}=3143~{\rm m}/{\rm s}$ and
$v_{r}=3937~{\rm m}/{\rm s}$\,.
The values on the vertical axes are to be multiplied by~$10^{25}$ and~$10^{22}$\,, on the left and right plots, respectively.}}
\label{fig:D2-vs-vr}
% from qray20161110.nb
\end{figure}

To solve numerically for~$v_{r}$\,, we let
\begin{equation}
XY-ST=:\mathrm{D}_{r2}=0\,,
\label{eq:Dr2}
\end{equation}
where solutions exist for particular values of~$C_{11}^u$\,,~$C_{44}^u$\,,~$\rho^u$\,,~$C_{11}^d$\,,~$C_{44}^d$\,,~$\rho^d$\,,~$Z$ and~$\omega$\,.
Note that arguments~$a$\,,~$a'$\,,~$b$\,,~$b'$ and expressions~(\ref{eq:rusurdsd}), for~$r^u$\,,~$r^d$\,,~$s^u$\,,~$s^d$\,, depend on~$v_{r}$; the arguments also depend on~$\kappa Z=\omega Z/v_r$\,.
Since the matrix includes frequency dependant terms, the quasi-Rayleigh waves---like Love waves but unlike classical Rayleigh waves---are dispersive.

Quasi-Rayleigh waves within the model discussed herein are reviewed by Ud\'{i}as~\cite[Section~10.4]{Udias1999} and Ben-Menahem and Singh~\cite[Section~3.6.5]{benmenahem}.
Unlike Ud\'{i}as~\cite{Udias1999}, we do not restrict Poisson's ratio in the layer and in the halfspace to be~$1/4$\,.
Furthermore, in our formulation, we found six corrections to Ud\'{i}as's formu{l\ae}, which are stated in \ref{sec:udias}.
Also, in a research paper, Fu~\cite{Fu1946} makes certain simplifying assumptions prior to calculations, which we do not. 
Such an exact approach allows for the examination of details regarding the forward problem and sets the stage for a further investigation.

\begin{figure}
\begin{center}
\includegraphics[scale=0.5]{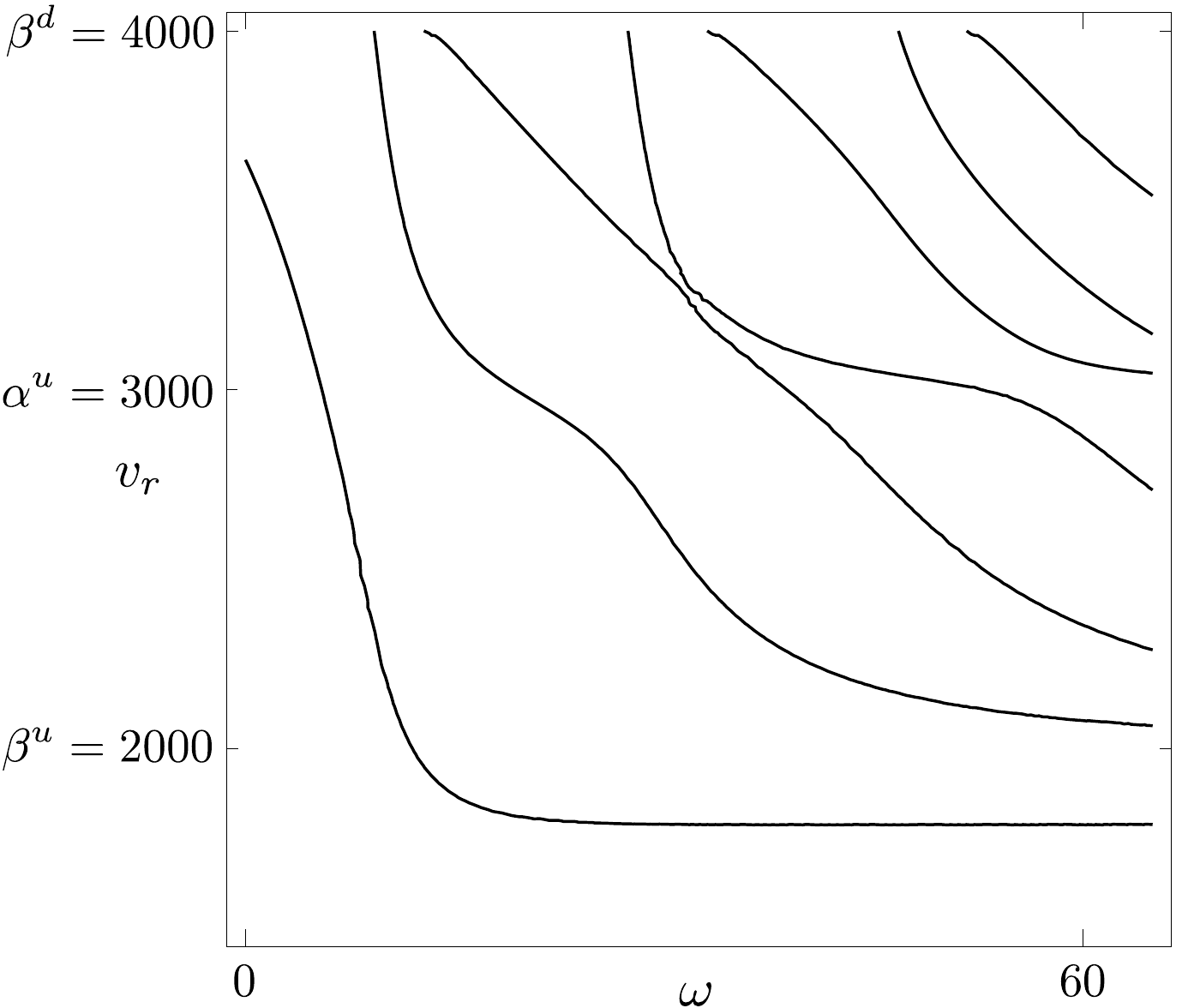}
\end{center}
\caption{\small{
The quasi-Rayleigh-wave dispersion curves,~$\mathrm{D}_{r2}=0$\,, defined in expression~(\ref{eq:Dr2}), as a function of speed,~$v_r$\,, and frequency,~$\omega$}}
% from 20160413qray-rusu.nb  or qray20161114.nb
\label{fig:qraycontour5}
\end{figure}

In Figure~\ref{fig:qraycontour5}, we plot the dispersion curves for expression~(\ref{eq:Dr2}), not for~$\mathrm{D}_r=0$\,, to avoid the trivial solutions,~$r^u=0$ and~$s^u=0$\,.
Also, in that manner, we avoid, for all frequencies, transitions between the real and imaginary domains, stated in expression~(\ref{eq:DrDomains}).
 
The existence of Love waves requires~$\beta^d > v_\ell  > \beta^u$\,, and of quasi-Rayleigh waves,~$\alpha^d > \beta^d > v_{r}$\,.
However, concerning propagation speeds, Ud\'{i}as~\cite{Udias1999} states that for the fundamental mode for high frequency,~$v_r<\beta^u$\,, but for higher modes,~$v_r>\beta^u$\,.
Therefore, it is not necessary that~$v_r>\beta^u$\,, except for higher modes.
Furthermore, the fundamental mode appears at all frequencies whereas higher modes have cutoff frequencies below which they do not appear.
If~$v_{r} <  \alpha^u$\,, then, from equations~(\ref{eq:QRGenSolAu}), (\ref{eq:QRuU1}) and~(\ref{eq:QRuU3}), which are found in~\ref{app:WaveEqnSols}, there is a partially exponential variation in the layer as opposed to a purely sinusoidal variation.
If~$\alpha^u \geq \beta^d$\,, the determinant of the coefficient matrix, as defined in expression~(\ref{eq:ourdet}), is purely imaginary but there remains a solution for~$v_{r}$ in~$\mathrm{D}_r=0$\,.

Let us examine, in an analogous manner to Figure~\ref{fig:love}, the numerical solutions of speed for the quasi-Rayleigh wave dispersion relation for high-frequency and low-frequency cases.
Herein, unlike for the Love wave, we need to introduce~$C_{11}^u=1.98\times 10^{10}~{\rm N}/{\rm m}^2$\,, which implies that~$\alpha^u=3000~{\rm m}/{\rm s}$\,, and~$C_{11}^d=10.985\times 10^{10}~{\rm N}/{\rm m}^2$\,, which implies that~$\alpha^d=6500~{\rm m}/{\rm s}$\,.

We depict in Figure~\ref{fig:D2-vs-vr} the high-frequency case, where~$\omega=60~{\rm s}^{-1}$\,, in the left plot and the low-frequency case, where~$\omega=15~{\rm s}^{-1}$\,, in the right plot.
Therein, we see that as~$\omega$ or~$Z$ increases, the number of solutions for~$v_{r}$ increases.
We observe that expression~(\ref{eq:ourdet}) is real or imaginary in the same manner as stated in expression~(\ref{eq:DrDomains}).
Yet,~$D_{r2}$, defined in expression~(\ref{eq:Dr2}), is real for the entire range of~$v_r$\,.

\begin{figure}
\begin{center}
\includegraphics[scale=0.5]{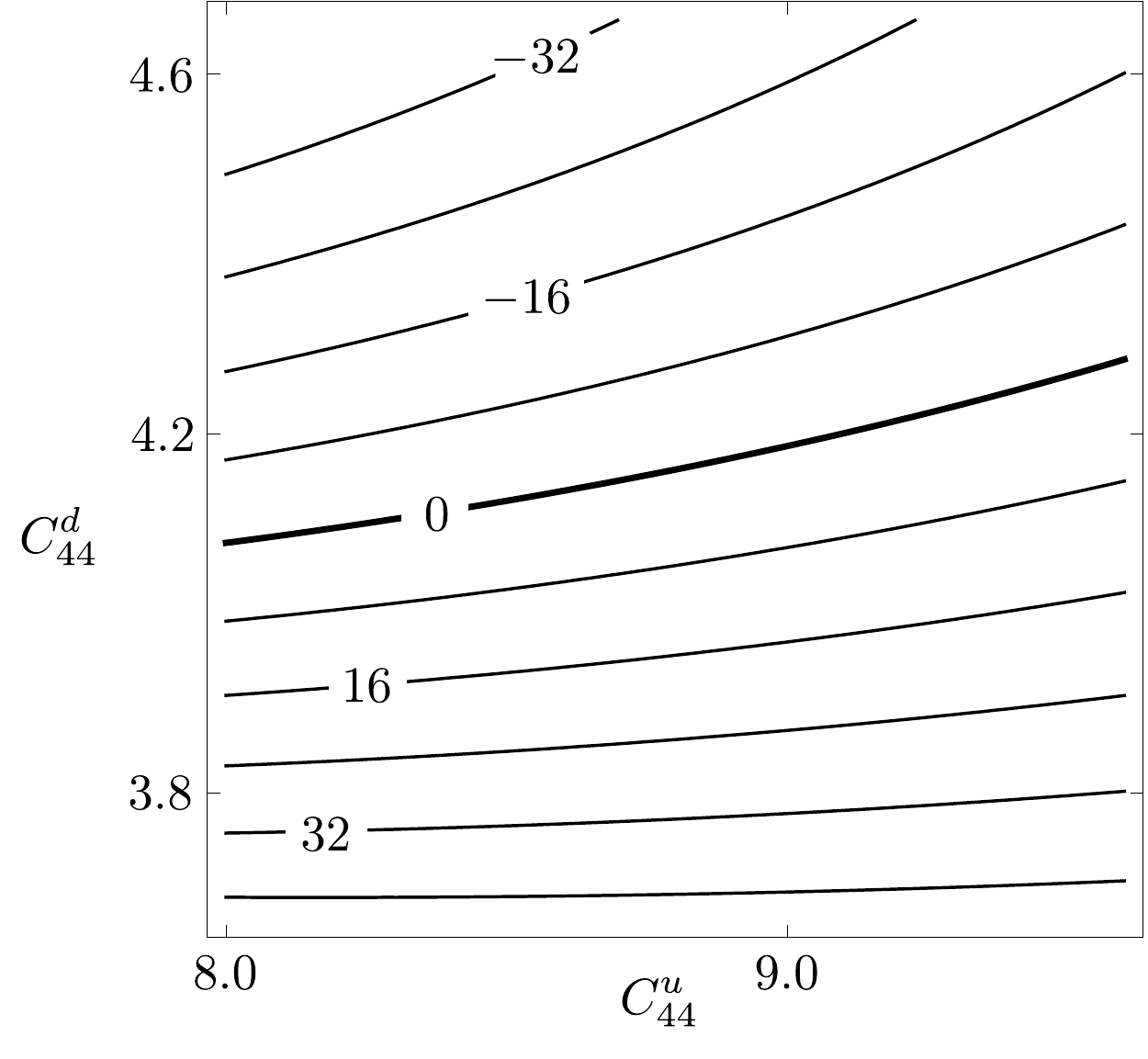}
\end{center}
\caption{\small{$\mathrm{D}_{r2}/10^{19}$\,, defined in expression~(\ref{eq:Dr2}),  as a function of the elasticity parameters,~$C_{44}^u$ and~$C_{44}^d$\,, for~$\omega=5~{\rm s}^{-1}$ and~$v_{r}=3069~{\rm m}/{\rm s}$\,.
 The values on the horizontal and vertical axes are to be multiplied by~$10^9$ and~$10^{10}$\,, respectively.  The sloping zero contour demonstrates sensitivity of the dispersion relation to both
$C_{44}^u$ and~$C_{44}^d$\,.}}
 % from qray20161114-2.nb
\label{fig:qraycontour1}
\end{figure}

Consider the case of~$v_r<\beta^u$ in Figure~\ref{fig:qraycontour5}, where the fundamental mode still has a solution for~$v_r$ for higher frequencies, but the higher modes do not.
For high frequency, the fundamental-mode speed asymptotically approaches the classical Rayleigh wave speed in the layer, which is~$0.89\beta^u$\,. 
In the limit---as~$\omega\rightarrow 0$\,---that fundamental mode---which unlike the higher modes has no low cutoff frequency---approaches the classical Rayleigh-wave speed in the halfspace, which is~$0.91\beta^d$\,.%from rayleightest.nb

\begin{figure}
\begin{center}
\hspace*{-0.1in}\includegraphics[scale=0.5]{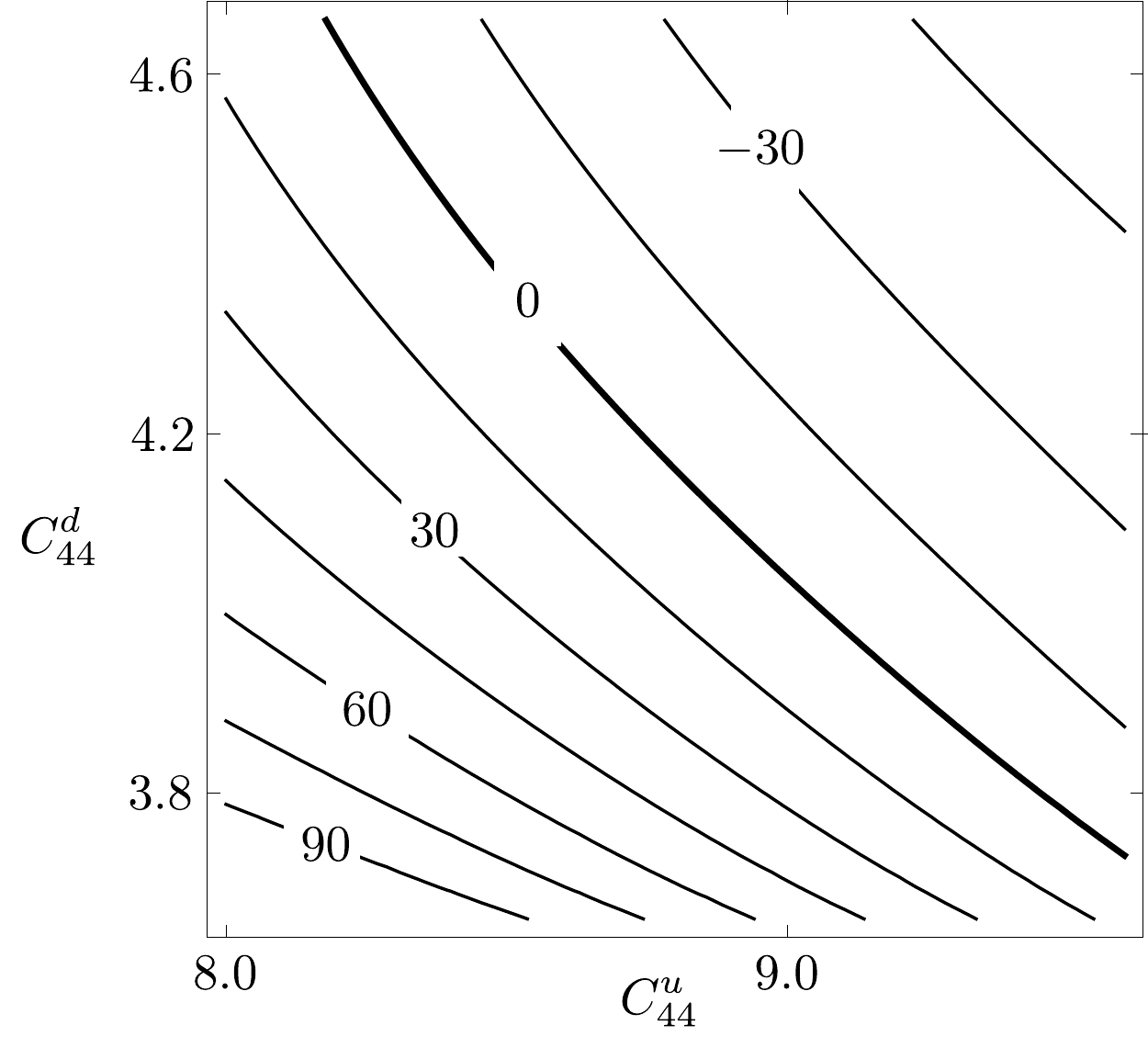}\hspace*{0.5in}\includegraphics[scale=0.5]{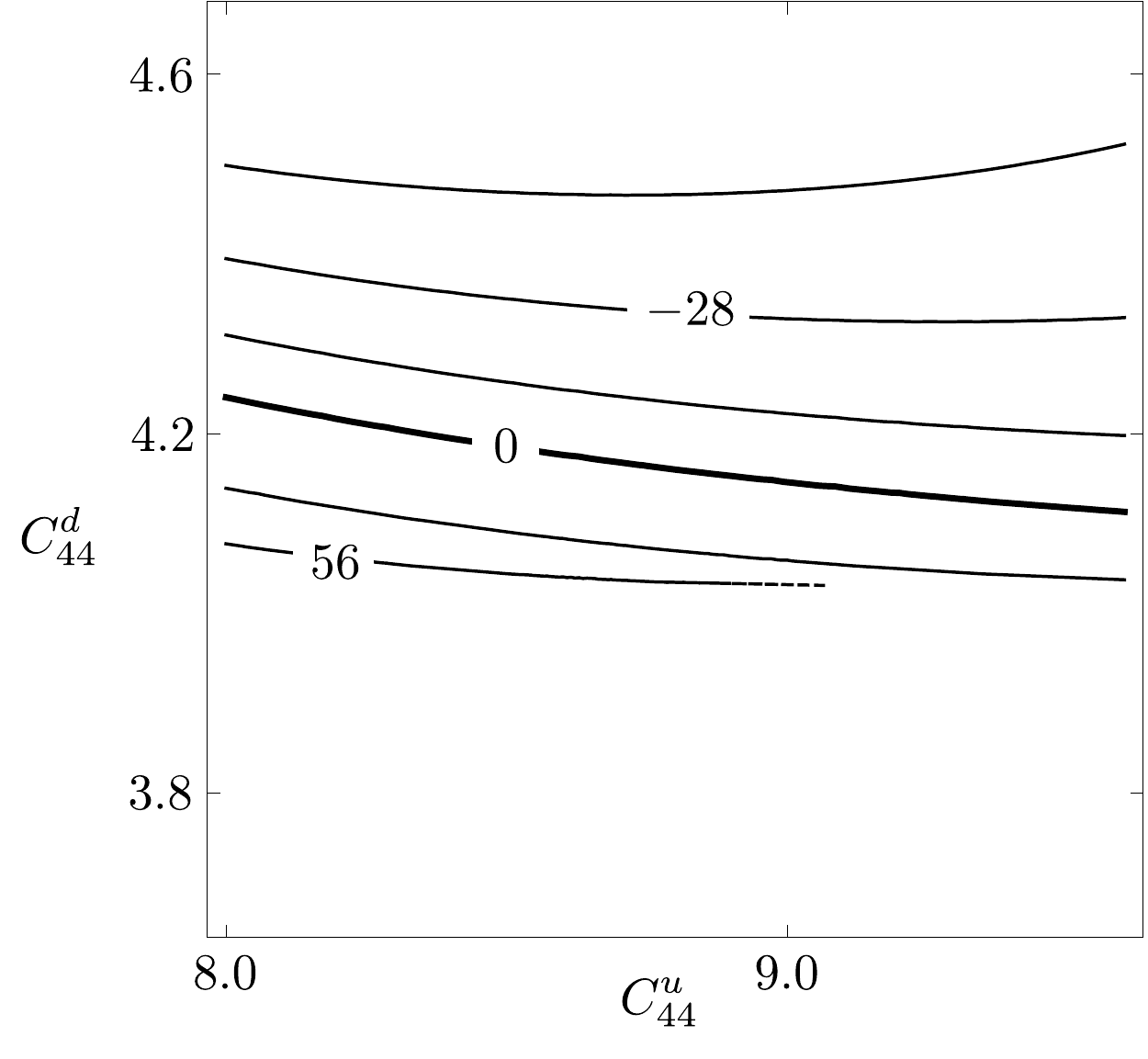}
%from qray20161114-2.nb
\end{center}
\caption{\small{
$\mathrm{D}_{r2}/10^{20}$\,, defined in expression~(\ref{eq:Dr2}),  as a function of the elasticity parameters,~$C_{44}^u$ and~$C_{44}^d$\,, 
for
$\omega=60~{\rm s}^{-1}$ and 
$v_r=3705~{\rm m}/{\rm s}$\,, on the left,
and for
$\omega=15~{\rm s}^{-1}$ and 
$v_r=3937~{\rm m}/{\rm s}$\,, on the right.
The values on the horizontal and vertical axes are to be multiplied by~$10^9$ and~$10^{10}$\,, respectively.
The zero contour on the left plot indicates sensitivity to both~$C_{44}^u$ and~$C_{44}^d$ but
the lower-frequency case, in the right plot, indicates more sensitivity to~$C_{44}^d$ than to~$C_{44}^u$\,.}}
\label{fig:qc-vhigh}
\end{figure}

\begin{figure}
\begin{center}
\hspace*{-0.1in}\includegraphics[scale=0.5]{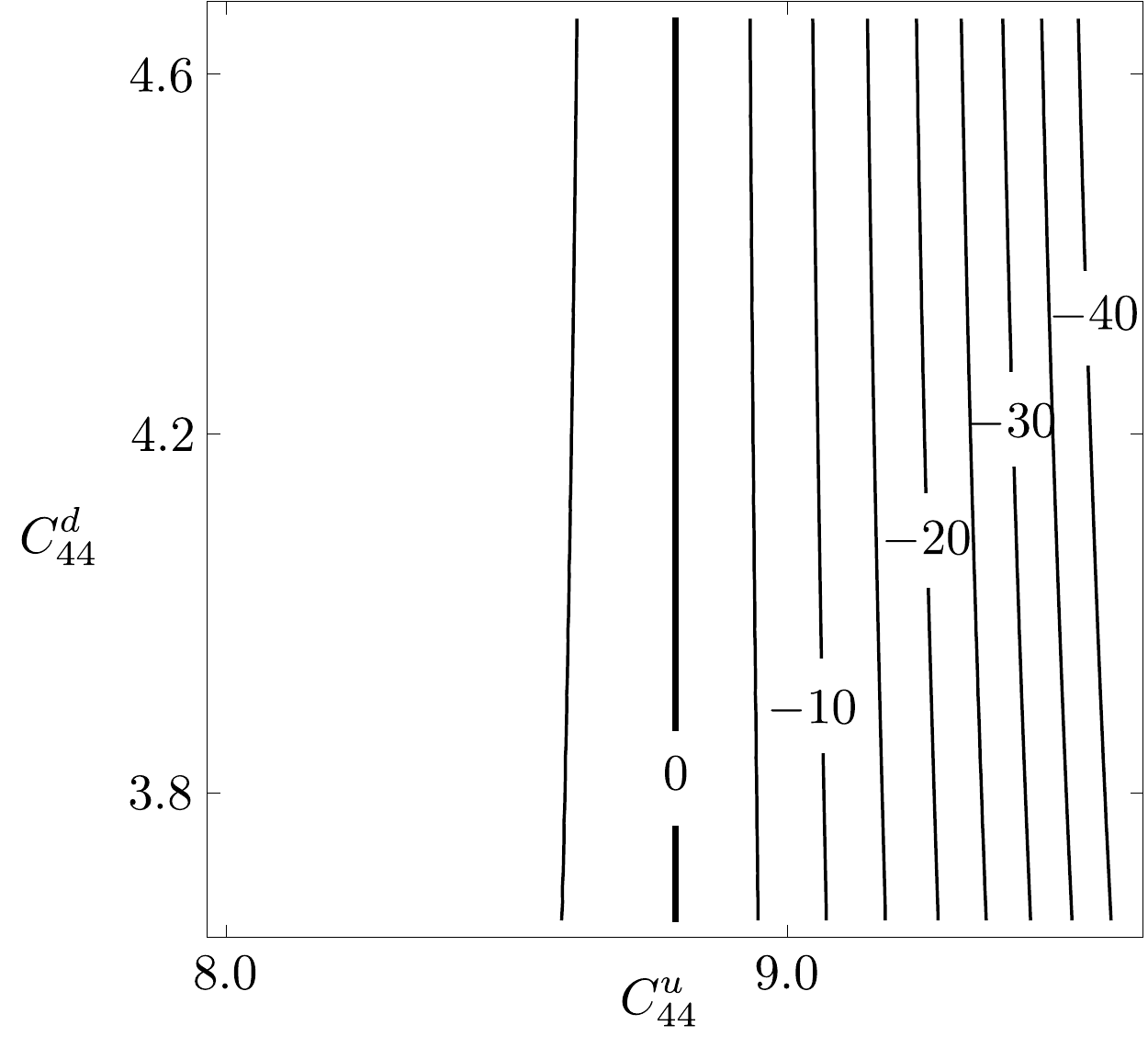}\hspace*{0.5in}\includegraphics[scale=0.5]{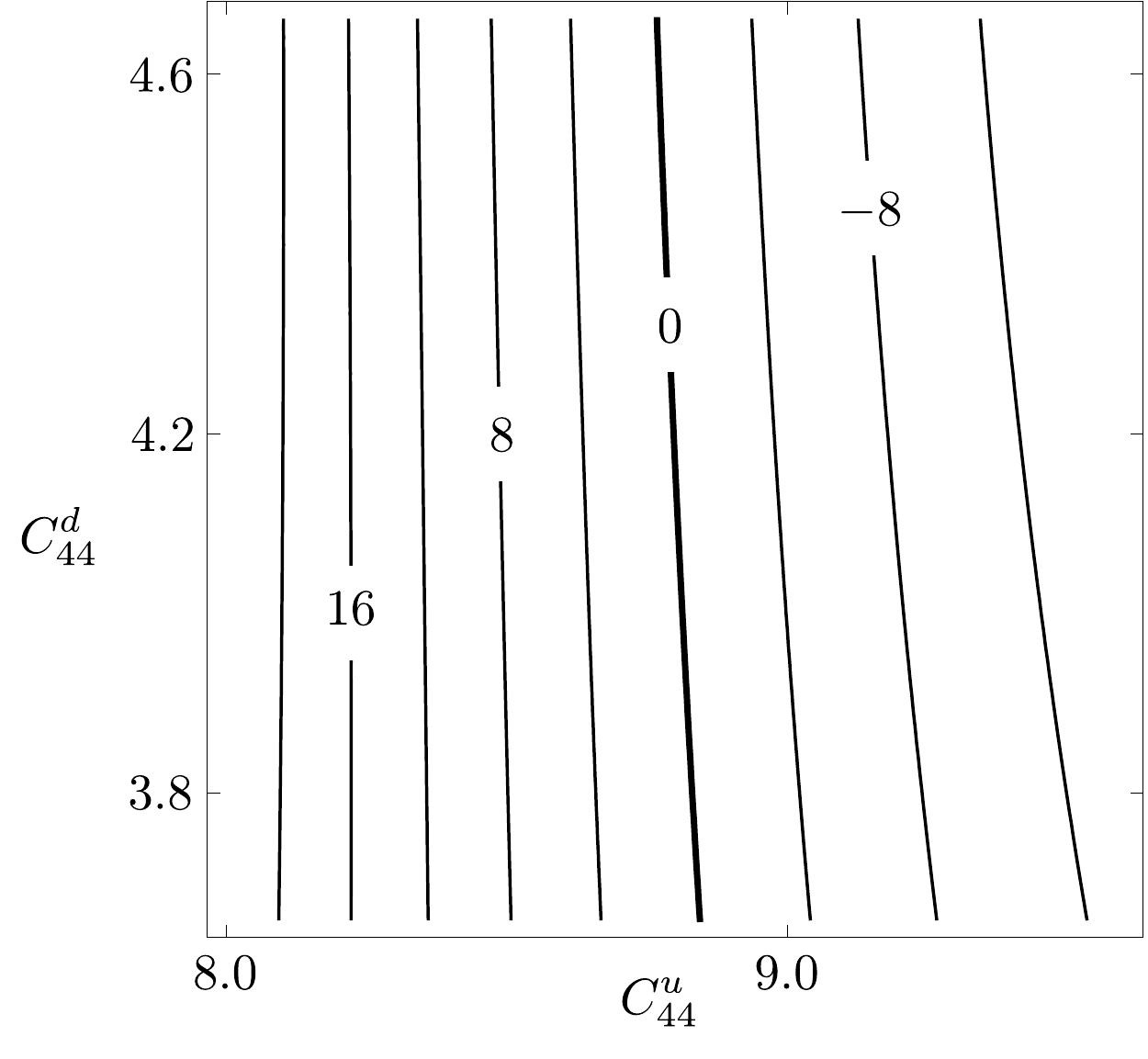}
%from qray20161114-2.nb
\end{center}
\caption{\small{
$\mathrm{D}_{r2}/10^{27}$\,, on the left, and~$\mathrm{D}_{r2}/10^{20}$\,, on the right, defined in expression~(\ref{eq:Dr2}),  as a function of the elasticity parameters,~$C_{44}^u$ and~$C_{44}^d$\,, where
$\omega=60~{\rm s}^{-1}$ and
$v_r=1786~{\rm m}/{\rm s}$\,,
and where
$\omega=15~{\rm s}^{-1}$ and
$v_r=1869~{\rm m}/{\rm s}$\,,
for left and right, respectively.
The values on the horizontal and vertical axes are to be multiplied by~$10^9$ and~$10^{10}$\,, respectively.
The near vertical contours in both plots indicate sensitivity to
$C_{44}^u$ and little sensitivity to~$C_{44}^d$\,.}}
\label{fig:qc-vlow}
\end{figure}

%%%%%%%%%%%%%%%%%%%%%%%%%%%%
\subsection{Sensitivity of dispersion relation}
%%%%%%%%%%%%%%%%%%%%%%%%%%%%
We wish to examine effects of elasticity parameters and layer thickness on the dispersion relation, for various frequencies and different modes.
To do so, we examine effects of these quantities on the value of~$\mathrm{D}_{r2}$\,, defined in expression~(\ref{eq:Dr2}).
Specifically, we examine~$\mathrm{D}_{r2}$ as a function of~$C_{44}^d$ and~$C_{44}^u$\,, for three distinct frequencies and for fixed values of~$v_r$\,, which correspond to particular modes.

\begin{figure}
\begin{center}
\includegraphics[scale=0.5]{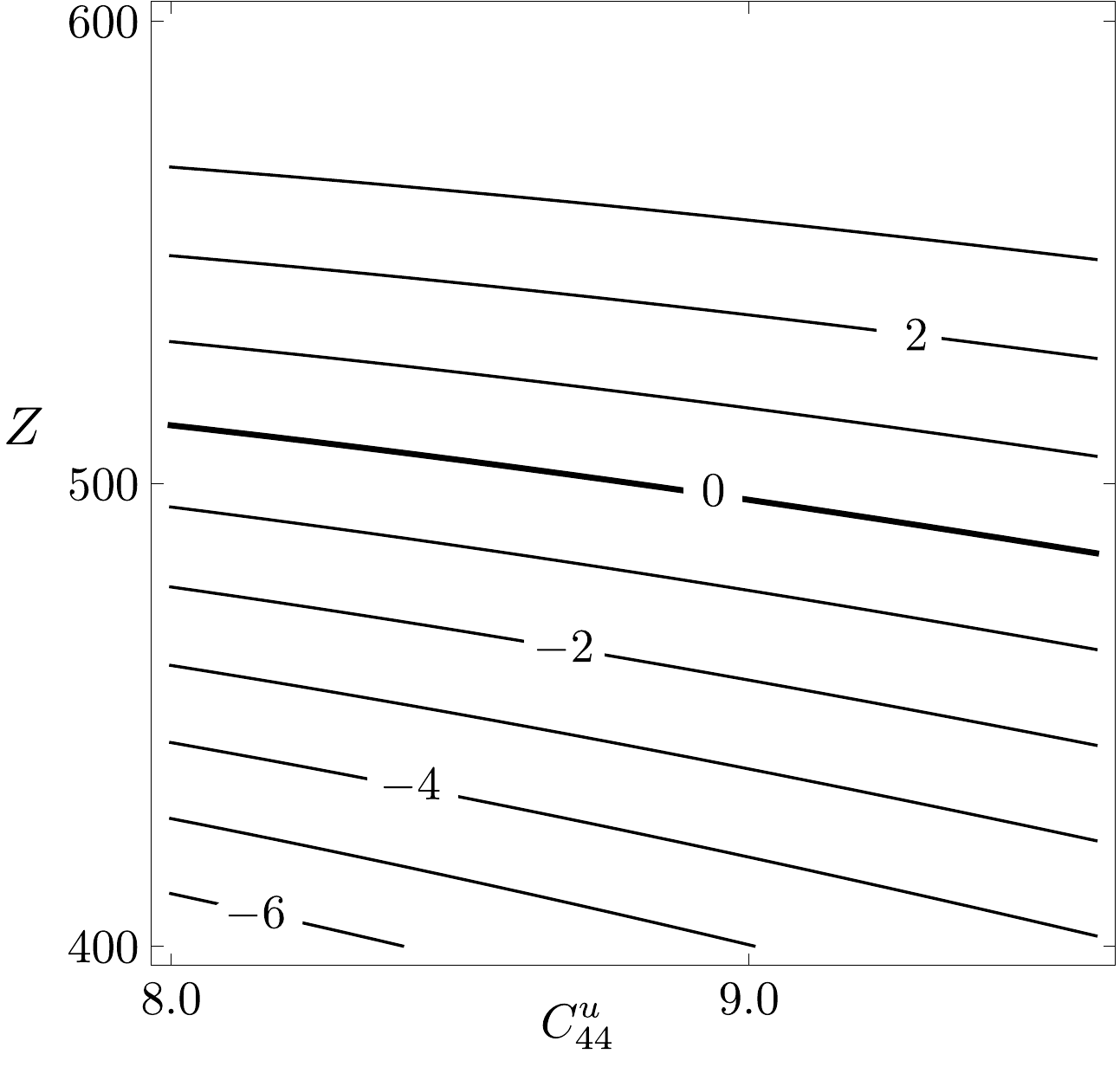}
\end{center}
\caption{\small{
$\mathrm{D}_{r2}/10^{20}$\,, defined in expression~(\ref{eq:Dr2}),  as a function of the elasticity parameter,~$C_{44}^u$\,, and layer thickness,~$Z$\,, for~$\omega=5~{\rm s}^{-1}$ and~$v_{r}=3069~{\rm m}/{\rm s}$\,.
The values on the horizontal axis are to be multiplied by~$10^9$\,. The sloping zero contour indicate sensitivity to both~$C_{44}^u$ and~$Z$\,.}}
 % from qray20161121.nb
\label{fig:qcZ-w5}
\end{figure}

\begin{figure}
\begin{center}
\hspace*{-0.1in}\includegraphics[scale=0.5]{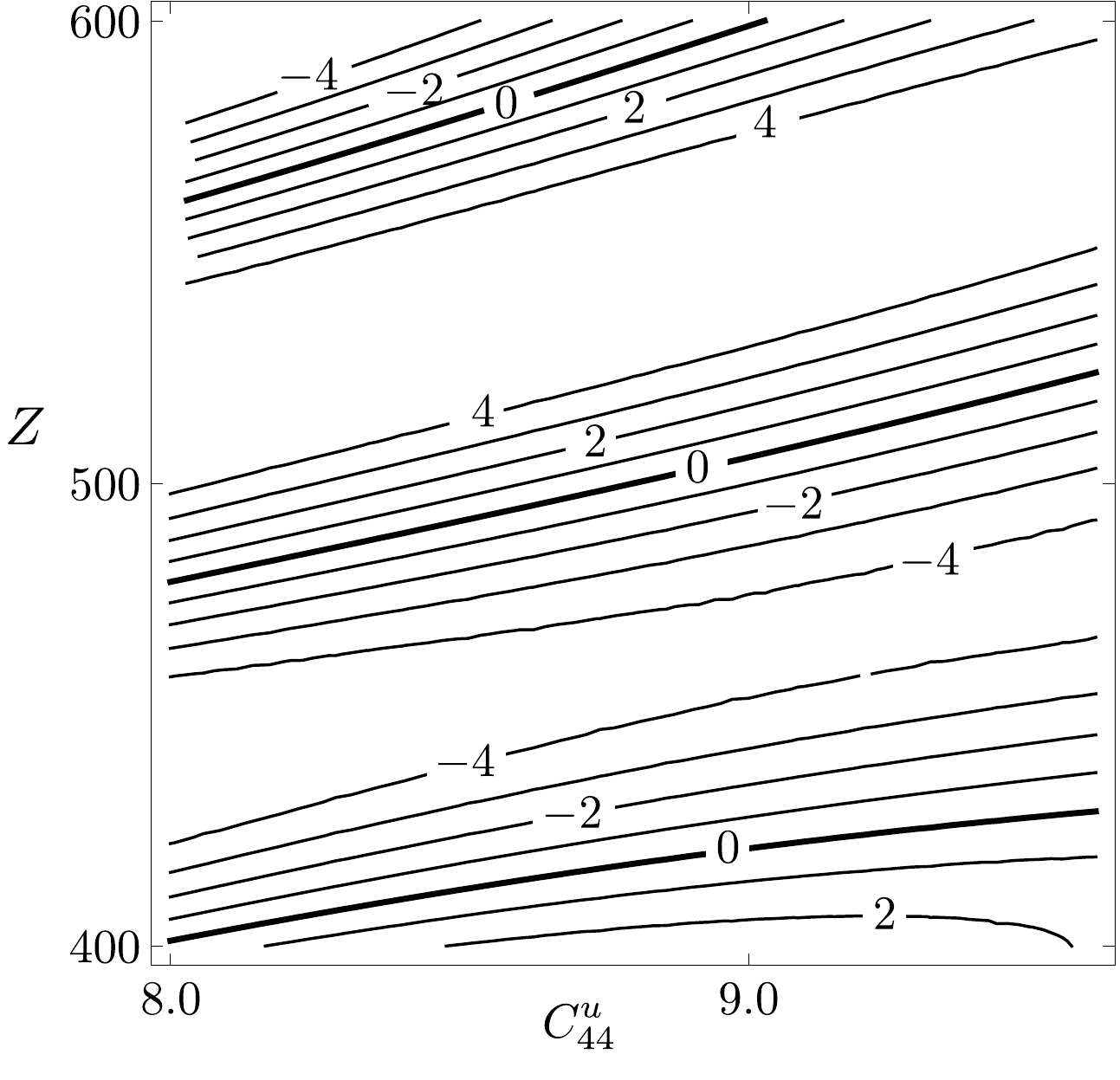}\hspace*{0.5in}\includegraphics[scale=0.5]{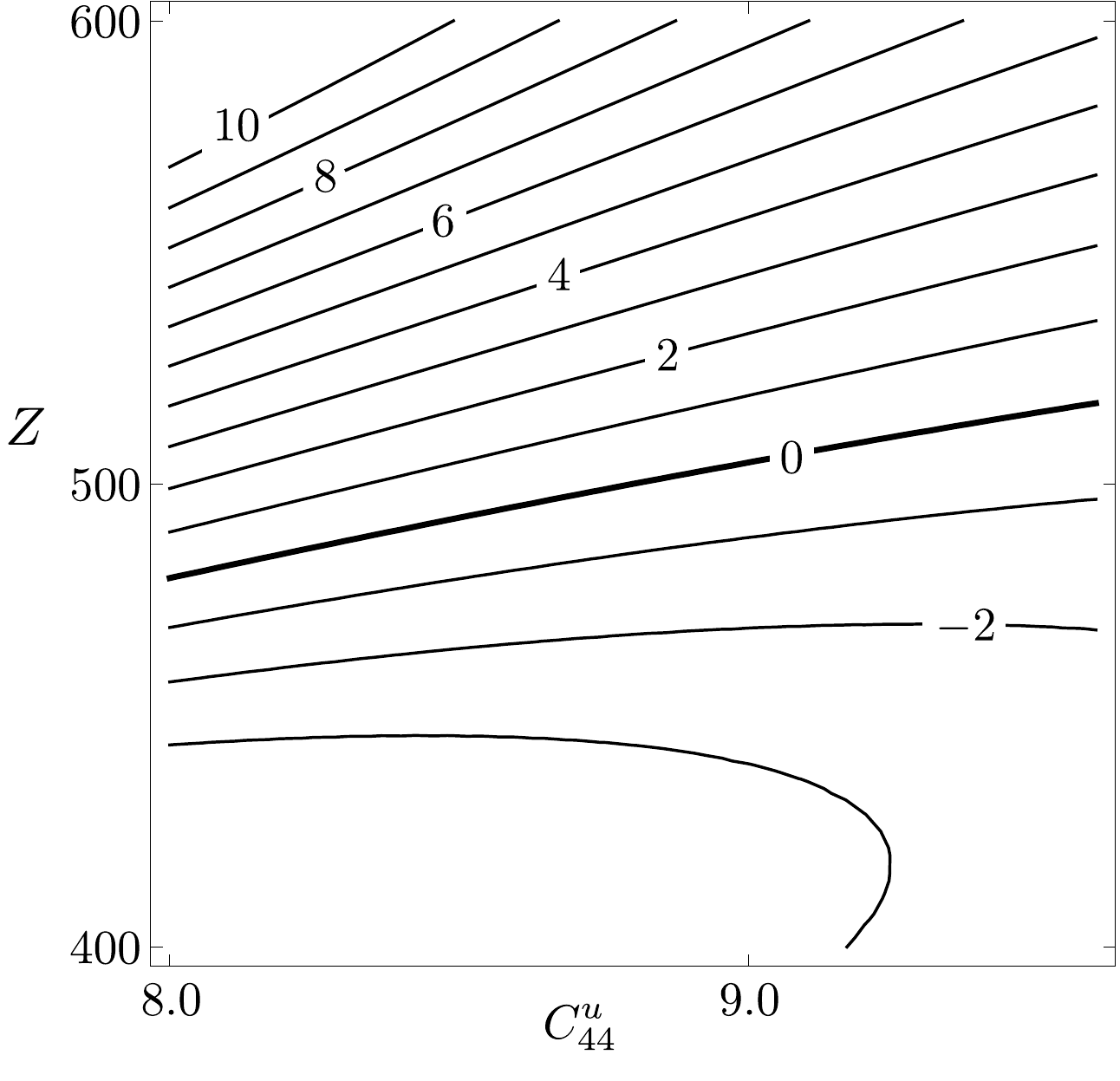}
%from qray20161121.nb
\end{center}
\caption{\small{
$\mathrm{D}_{r2}/10^{21}$\,, defined in expression~(\ref{eq:Dr2}),  as a function of the elasticity parameters,~$C_{44}^u$\,, and layer thickness,~$Z$\,, 
for
$\omega=60~{\rm s}^{-1}$ and 
$v_r=3705~{\rm m}/{\rm s}$\,, on the left,
and for
$\omega=15~{\rm s}^{-1}$ and 
$v_r=3937~{\rm m}/{\rm s}$\,, on the right.
The values on the horizontal axis is to be multiplied by~$10^9$\,.
The sloping
 contours indicate sensitivity to both~$C_{44}^u$ and~$Z$\,.}}
\label{fig:qcZ-vhigh}
\end{figure}

\begin{figure}
\begin{center}
\hspace*{-0.1in}\includegraphics[scale=0.5]{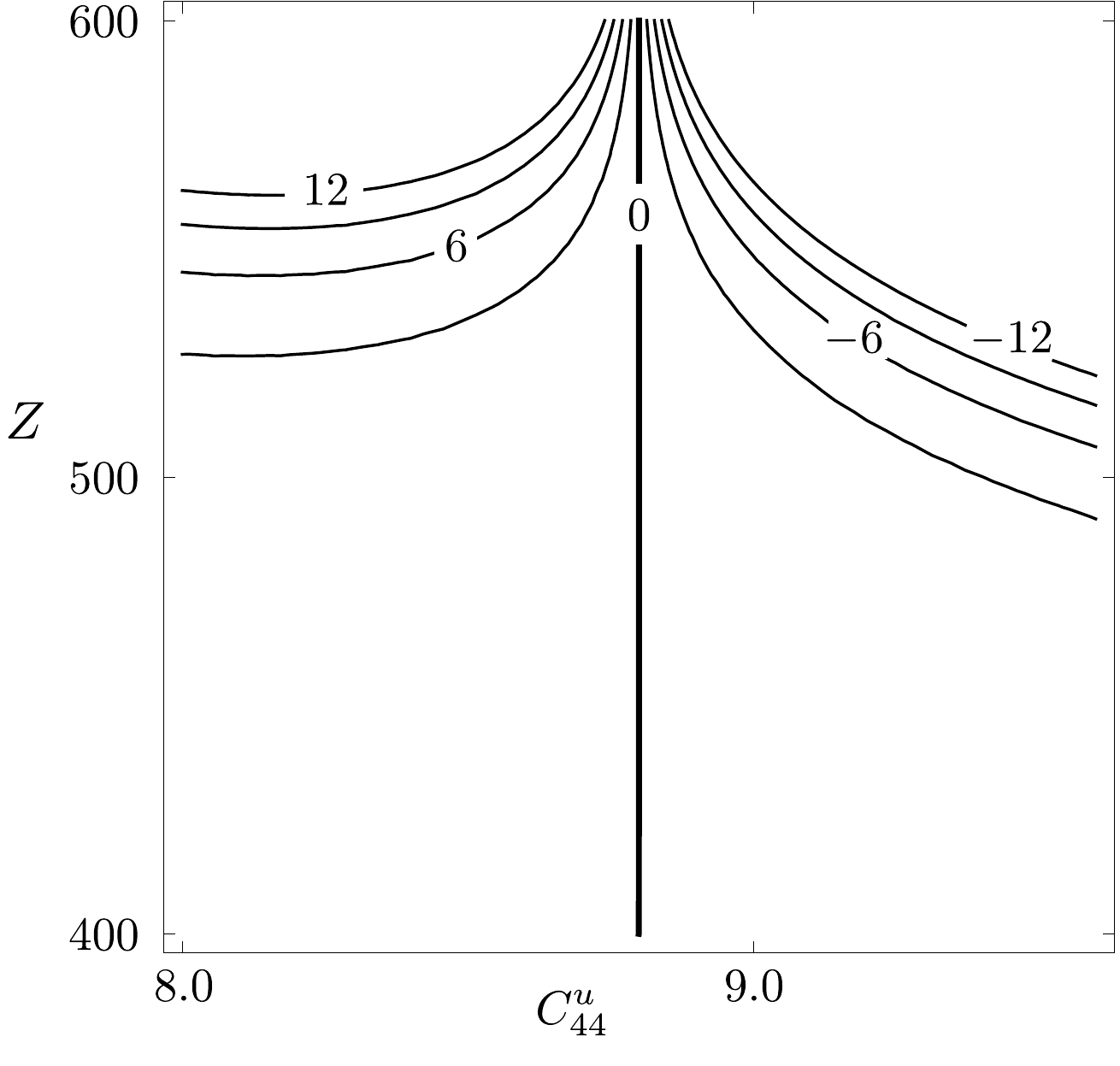}\hspace*{0.5in}\includegraphics[scale=0.5]{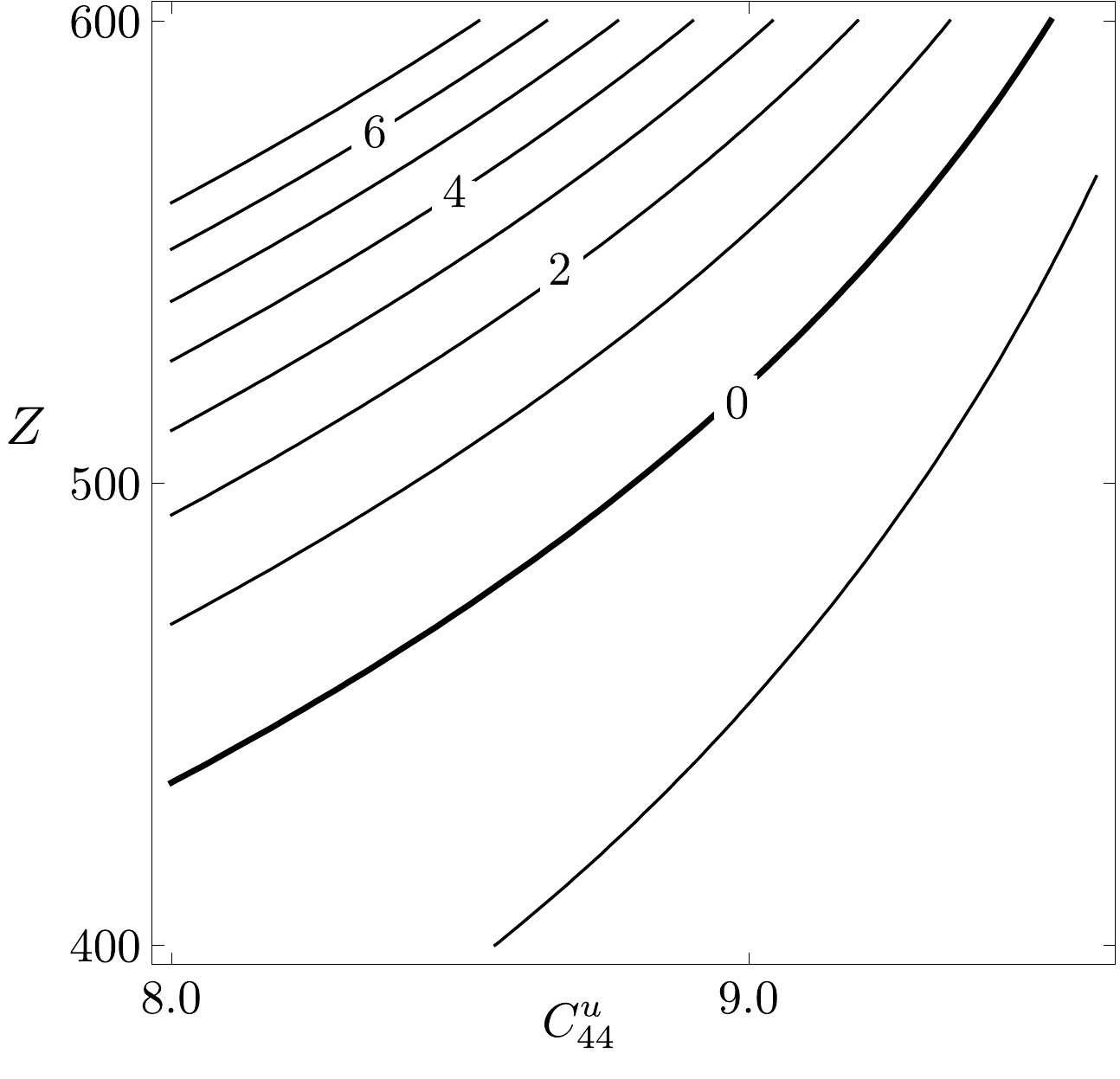}
%from qray20161121.nb
\end{center}
\caption{\small{
$\mathrm{D}_{r2}/10^{28}$\,, on the left, and~$\mathrm{D}_{r2}/10^{21}$\,, defined in expression~(\ref{eq:Dr2}), on the right, as a function of the elasticity parameter,~$C_{44}^u$\,, and layer thickness,~$Z$\,, where
$\omega=60~{\rm s}^{-1}$ and
$v_r=1786~{\rm m}/{\rm s}$\,,
and where
$\omega=15~{\rm s}^{-1}$ and
$v_r=1869~{\rm m}/{\rm s}$\,,
for the left and right plots, respectively.
The values on the horizontal axis is to be multiplied by~$10^9$\,.
The sloping zero contour in the lower frequency plot indicates sensitivity to both~$C_{44}^u$ and~$Z$ while the vertical zero contour in the  higher frequency plot indicates sensitivity to~$C_{44}^u$ but no sensitivity to~$Z$\,.}}
\label{fig:qcZ-vlow}
\end{figure}

The contour maps of Figures~\ref{fig:qc-vhigh},~\ref{fig:qc-vlow},~\ref{fig:qcZ-vhigh} and~\ref{fig:qcZ-vlow} illustrate the sensitivity of the dispersion relation by the slope of the zero contour line for quasi-Rayleigh waves.
These plots, and their interpretation, are analogous to Figures~\ref{fig:lc-vhigh}--\ref{fig:lcZ-vlow}.

Let us examine the numerical solutions for the quasi-Rayleigh wave dispersion relation for the high-frequency and low-frequency cases.
We use these solutions to investigate the relative sensitivities of the quasi-Rayleigh wave to the elasticity parameter in the upper layer and in the lower halfspace, as well as to the elasticity parameter in the upper layer and to the layer thickness.

We begin by considering the layer and halfspace elasticity parameters.
For the high-speed case, we consider the seventh root of the left plot of Figure~\ref{fig:D2-vs-vr}, which is~$v_{r}=3705~{\rm m}/{\rm s}$\,, and the third root of the right plot of Figure~\ref{fig:D2-vs-vr}, which is~$v_{r}=3937~{\rm m}/{\rm s}$\,.
The left and right plots of Figure~\ref{fig:qc-vhigh} are the corresponding contour plots of~$\mathrm{D}_{r2}/10^{20}$ with varying~$C_{44}^u$ and~$C_{44}^d$\,.
In both cases,~$\mathrm{D}_{r2}$ is sensitive to~$C_{44}^u$ and~$C_{44}^d$\,.
However, the right plot of Figure~\ref{fig:qc-vhigh}, which depicts a lower frequency and high speed, indicates a greater sensitivity to~$C_{44}^d$\,.

Then, we consider Figure~\ref{fig:qraycontour1}, which is the corresponding plot of~$\mathrm{D}_{r2}/10^{19}$\,, whose frequency and speed is lower than the right plot of Figure~\ref{fig:qc-vhigh}.
In this case, we observe that the determinant is also sensitive to both~$C_{44}^u$ and~$C_{44}^d$\,.

For the low-speed case, we observe different sensitivities.
Let us examine the first root of the left plot of Figure~\ref{fig:D2-vs-vr}, which is~$v_{r}=1786~{\rm m}/{\rm s}$\,, and the first root of the right plot of Figure~\ref{fig:D2-vs-vr}, which is~$v_{r}=1869~{\rm m}/{\rm s}$\,.
Following the corresponding plot of Figure~\ref{fig:qc-vlow}, we see that there are near vertical zero lines.
This indicates that there is a greater sensitivity~$C_{44}^u$ but lower sensitivity to~$C_{44}^d$ for both frequencies.

Next, we consider the layer elasticity parameter and the layer thickness.
Using the high-speed roots of expression~(\ref{eq:Dr2}), we observe sensitivity to both~$C_{44}^u$ and~$Z$\,, in the high-frequency and low-frequency cases, which are depicted in Figure~\ref{fig:qcZ-vhigh}.
Using low-speed roots in Figure~\ref{fig:qcZ-vlow}, we observe sensitivity in the right plot to both~$C_{44}^u$ and~$Z$.
However, the left plot of Figure~\ref{fig:qcZ-vlow} has no sensitivity to~$Z$ since, regardless of~$Z$\,, the fundamental-mode speed is asymptotic to the quasi-Rayleigh wave speed in the layer for higher frequencies.

We consider Figure~\ref{fig:qcZ-w5} as well, which is the corresponding plot of~$\mathrm{D}_{r2}/10^{20}$\,, whose frequency and speed is lower than the right plot of Figure~\ref{fig:qcZ-vhigh}.
In this case, we observe that the determinant is sensitive to both~$C_{44}^u$ and~$C_{44}^d$\,.

Additionally, Figure~\ref{fig:qcZ-vhigh} is similar to Figure~\ref{fig:lcZ-vhigh}.
The periodicity that shows up on the left plot of Figure~\ref{fig:qcZ-vhigh} is due to the periodicity in~$D_{r2}$\,.  
Furthermore, the left plot of Figure~\ref{fig:qcZ-vlow} is distinct from the left plot of Figure~\ref{fig:lcZ-vlow}.
The distinction is due to differences in asymptotic behaviour of the fundamental mode at high frequencies for the dispersion curves of Love and quasi-Rayleigh waves.
%%%%%%%%%%%%%%%%%%%%%%%%%%%%
\section{Conclusions}
%%%%%%%%%%%%%%%%%%%%%%%%%%%%
\subsection{Background examination}
%%%%%%%%%%%%%%%%%%%%%%%%%%%%
In this paper, we study the sensitivity of Love waves and quasi-Rayleigh waves to model parameters based on their exact dispersion relations.
Both are guided waves in an elastic layer constrained by a vacuum and an elastic halfspace.
According to presented computations, and as illustrated in Figures~\ref{fig:love3x3contourw5-2} and~\ref{fig:qraycontour5}, the dispersion relations for the Love and quasi-Rayleigh waves imply that their speeds vary with frequency and differ from one another.
Thus, their arrival times should be distinct on a seismic record, which---together with the fact that their polarizations are orthogonal to one other---makes them independent sources of information to infer model parameters.

The speeds of both Love and quasi-Rayleigh waves, whose roots are depicted in Figures~\ref{fig:love} and~\ref{fig:D2-vs-vr}, respectively, are obtained from their dispersion relations.
In the high-frequency case, the fundamental Love-wave mode has a speed that is slightly greater than the~$S$-wave speed in the layer.
In the low-frequency case, its speed is slightly lower than the~$S$-wave speed in the halfspace.
The highest-mode speeds of both the Love wave and the quasi-Rayleigh wave are less than the~$S$-wave speed in the halfspace.

The dispersion curves for Love waves and quasi-Rayleigh waves are given in Figure~\ref{fig:love3x3contourw5-2} and Figure~\ref{fig:qraycontour5}, respectively.
In the latter figure---and in agreement with Figure~10.14 of Ud\'{i}as~\cite{Udias1999}---the fundamental mode has all frequencies, which means that it has no cutoff frequency.
For high frequency, the fundamental-mode speed asymptotically approaches the classical Rayleigh-wave speed in the layer, and---in the limit as~$\omega\rightarrow 0$\,---that speed approaches the classical Rayleigh-wave speed in the halfspace.  
 
We review the dispersion relation for Love waves and provide the details in appendices of the derivation of the dispersion relation for quasi-Rayleigh waves, including details of the expansion of the~$6\times 6$ matrix and its determinant.
We compare our results to several past studies, including the one of Love~\cite{Love1911}, which assumes incompressibility, and the one of Ud\'{i}as~\cite{Udias1999}, in which we found errors in the equations, though not in the dispersion-curve plots.
Unlike Love~\cite{Love1911}, who assumes incompressibility, Ud\'{i}as~\cite{Udias1999}, who assumes a Poisson's ratio of~$1/4$\,, and Fu~\cite{Fu1946}, who studies limiting cases, we do not make any simplifying assumptions prior to calculations for the study of sensitivity.

In the case of quasi-Rayleigh waves, and in the context of the~$6\times 6$ determinant, we show that the solutions~$r^u=0$ and~$s^u=0$ have zero displacements, and hence, can be considered trivial solutions.
%%%%%%%%%%%%%%%%%%%%%%%%%%%%
\subsection{Sensitivity results}
\label{sec:SensitivityResults}
%%%%%%%%%%%%%%%%%%%%%%%%%%%%
Our results demonstrate the sensitivity of the Love wave and quasi-Rayleigh wave dispersion relations to elasticity parameters and layer thickness.
We conclude that the fundamental mode is mainly sensitive to the upper layer properties while higher modes are sensitive to both the upper layer and lower halfspace properties.
Within each mode the lower frequencies are more sensitive to the lower halfspace than are higher frequencies.

We do not consider sensitivity to~$C_{11}$ since we remain within the elasticity parameters common for both the Love and quasi-Rayleigh waves, nor do we examine the sensitivity to~$\rho$\,.
Sensitivities to~$C_{11}$ and~$\rho$ are obtainable by similar procedures.
However, according to Lucena and Taioli~\cite{Lucena}, the dispersion curves are sensitive to neither of them.

In the case of Love waves, we formulate and examine the absolute value of a dimensionless sensitivity coefficient given in terms of partial derivatives of the Love-wave dispersion relation with respect to dimensionless variables,~$\nu$ and~$\theta$\,.
From these results, we perform an analysis to deduce the optimum frequency,~$\omega_0$\,, to obtain~$Z$ from a given mode.

Our results, in particular, Figures~\ref{fig:love3x3contourw5-2},~\ref{fig:sensitivity2},~\ref{fig:sensitivity3} and~\ref{fig:qraycontour5}, allow us to infer data-acquisition information.
This can be achieved by considering frequencies required to increase the number of modes.
Using this information, we may calibrate sources and receivers to record different modes.
For example, at~$\omega=30~{\rm s}^{-1}$\,, we can see two Love-wave modes and four quasi-Rayleigh-wave modes.
%%%%%%%%%%%%%%%%%%%%%%%%%%%%%%%
\section{Future work}
\label{sec:FutureWork}
%%%%%%%%%%%%%%%%%%%%%%%%%%%%%%%
Given dispersion relations of the Love and the quasi-Rayleigh waves, we expect to invert the measurements of speed for elasticity parameters and mass densities of both the layer and the halfspace, as well as for the layer thickness.
Explicit expressions presented in this paper allow us to formulate the inverse problem and examine the sensitivity of its solution.  
In particular, the presence of two types of waves lends itself to the formulation of a joint inversion, which exploits the redundancy of information.  
The presented study of the Love and quasi-Rayleigh wave sensitivities allows us to gain an insight into their combination, which appears explicitly in such an inversion.

In further studies, we could formulate dispersion relations for the case of an anisotropic layer and an anisotropic halfspace.
Such a formulation would require modified boundary conditions and equations of motion.
Further insights into such issues are given by Babich and Kiselev~\cite{Babich}.
%%%%%%%%%%%%%%%%%%%%%%%%%%%%%%%
\section*{Acknowledgments}
%%%%%%%%%%%%%%%%%%%%%%%%%%%%
We wish to acknowledge discussions with Adrian Bogacz, Tomasz Danek and Michael Rochester.
We also acknowledge the graphical support of Elena Patarini.  Also, we wish to
acknowledge insightful suggestions of Paul Martin, as the editor, and
the two anonymous reviewers.
This research was performed in the context of The Geomechanics Project supported by Husky Energy. 
Also, this research was partially supported by the Natural Sciences and Engineering Research Council of Canada, grant 238416-2013.
%\nocite{*}
\bibliographystyle{unsrt}
\bibliography{qrayandlovebib}

\begin{thebibliography}{1}

\bibitem{Lucena}
R.~F.~de Lucena and F.~Taioli.
\newblock Rayleigh wave modeling: {A} study of dispersion curve sensitivity and
  methodology for calculating an initial model to be included in an inversion
  algorithm.
\newblock {\em Journal of Applied Geophysics}, 108:140--151, 2014.

\bibitem{SlawinskiWS2016}
M.~A. Slawinski.
\newblock {\em Waves and rays in seismology: {A}nswers to unasked questions}.
\newblock World Scientific, 2016.

\bibitem{Udias1999}
A.~Ud\'{i}as.
\newblock {\em Principles of seismology}.
\newblock Cambridge University Press, 1999.

\bibitem{Novotny}
O.~Novotn\'{y}.
\newblock Methods of computing the partial derivatives of dispersion curves.
\newblock {\em Pageoph}, 114:765--774, 1976.

\bibitem{benmenahem}
A.~Ben-Menahem and S.~J. Singh.
\newblock {\em Seismic waves and sources}.
\newblock Dover, Mineola, N.Y., 2nd edition, 2000.

\bibitem{Fu1946}
C.~Y. Fu.
\newblock Studies on seismic waves: I{I}. {R}ayleigh waves in a superficial
  layer.
\newblock {\em Geophysics}, 11(1):10--23, 1946.

\bibitem{Love1911}
A.~E.~H. Love.
\newblock {\em Some problems of geodynamics}.
\newblock Cambridge University Press, 1911.

\bibitem{Babich}
V.~M. Babich and A.~P. Kiselev.
\newblock {\em Elastic waves: High-frequency theory (in Russian)}.
\newblock BHV St. Petersburg, 2014.

\bibitem{Lee1932}
A.~W. Lee.
\newblock The effect of geological structure upon microseismic disturbance.
\newblock {\em Geophysical Journal International}, 3:85--105, 1932.

\end{thebibliography}
%%%%%%%%%%%%%%%%%%%%%%%%%%%%%%%%
\setcounter{section}{0}
\renewcommand{\thesection}{Appendix~\Alph{section}}
%%%%%%%%%%%%%%%%%%%%%%%%%%%%%%%%
\section{Formulation of quasi-Rayleigh waves}
\label{app:qRwaves}
%%%%%%%%%%%%%%%%%%%%%%%%%%%%
\subsection{Material properties and wave equations}
%%%%%%%%%%%%%%%%%%%%%%%%%%%%
\footnote{The formulation in this section is similar to that of Ud\'{i}as~\cite[Section~10.4]{Udias1999} except that we set~$x_3$ to be positive downwards with the free surface at~$x_3=0$ and interface at~$x_3=Z$\,, whereas Ud\'{i}as~\cite{Udias1999} sets~$x_3$ to be positive upwards with the free surface at~$x_3=H$ and interface at~$x_3=0$\,.
More importantly, unlike Ud\'{i}as~\cite{Udias1999}, we do not restrict Poisson's ratio in the layer and in the halfspace to~$1/4$\,.  Also we found some errors
in Ud\'{i}as's equations.}
Using the Helmholtz decomposition theorem, we express the displacements as
\begin{equation}
\label{eq:QRpotentials}
u^\mu=\nabla{\mathcal P}^\mu+\nabla\times\mbox{\boldmath${\mathcal S}$}^\mu\,,\quad u_1^\mu=\frac{\partial {\mathcal P}^\mu}{\partial x_1}
-\frac{\partial {\mathcal S}^\mu}{\partial x_3}\,,\quad u_3^\mu=\frac{\partial {\mathcal P}^\mu}{\partial x_3}+
\frac{\partial {\mathcal S}^\mu}{\partial x_1}\,,\quad u_2^\mu=0\,,\,\,\mu=u,d,
\end{equation}
where we use the gauge condition outlined in Slawinski~\cite[Section~6.2]{SlawinskiWS2016} and set~${\mathcal S}_2^\mu={\mathcal S}^\mu$\,, for brevity.
$\mathcal P^\mu$ denotes the scalar potential and~$\mbox{\boldmath${\mathcal S}$}^\mu=[{\mathcal S}_1^\mu,{\mathcal S}_2^\mu,{\mathcal S}_3^\mu]$ denotes the vector potential, which herein is~$\mbox{\boldmath${\mathcal S}$}^\mu=[0,{\mathcal S}^\mu,0]$\,.
Potentials allow us to consider the coupling between the~$P$ and~$SV$ waves.
The pertinent wave equations are
\begin{equation}
\label{eq:QRWaveEqns}
\nabla^2 {\mathcal P}^\mu-\frac{1}{(\alpha^\mu)^2}\frac{\partial^2 {\mathcal P}^\mu}{\partial t^2}=0\,\,,\,\,\,
\nabla^2 {\mathcal S}^\mu-\frac{1}{(\beta^\mu)^2}\frac{\partial^2 {\mathcal S}^\mu}{\partial t^2}=0\,,\,\,\,\mu=u,d,
\end{equation}
which correspond to the~$P$ waves and~$SV$ waves, respectively.
%%%%%%%%%%%%%%%%%%%%%%%%%%%%
\subsection{Solutions of wave equations}
\label{app:WaveEqnSols}
%%%%%%%%%%%%%%%%%%%%%%%%%%%%
Let the trial solutions for the corresponding wave equations be
\begin{equation}
\label{eq:QRAnsatzP}
{\mathcal P}^\mu=A^\mu(x_3)\exp(\iota(\kappa x_1-\omega t))\,,\quad\mu=u,d\,,
\end{equation}
\begin{equation}
\label{eq:QRAnsatzS}
{\mathcal S}^\mu=B^\mu(x_3)\exp(\iota(\kappa x_1-\omega t))\,,\quad\mu=u,d\,.
\end{equation}
Inserting solutions~(\ref{eq:QRAnsatzP}) and (\ref{eq:QRAnsatzS}) into equations~(\ref{eq:QRWaveEqns}) leads to
\begin{equation*}
\frac{\,{\rm d}^2 A^\mu}{\,{\rm d}x_3^2}+\left(\frac{\omega^2}{(\alpha^\mu)^2}-\kappa^2\right)A^\mu=0\,,\,\,\,
\frac{\,{\rm d}^2 B^\mu}{\,{\rm d}x_3^2}+\left(\frac{\omega^2}{(\beta^\mu)^2}-\kappa^2\right)B^\mu=0\,,\,\,\,\mu=u,d\,,
\end{equation*}
which are ordinary differential equations for amplitudes~$A^\mu$ and~$B^\mu$\,. 
Similarly to the derivation of Love waves, we require displacements to decay  within the halfspace.
Expressions~(\ref{eq:QRpotentials}), which denote displacements, entail
\begin{equation*}
\kappa^2-\omega^2/(\alpha^d)^2 > 0\quad{\rm and}\quad\kappa^2-\omega^2/(\beta^d)^2 > 0\,.
\end{equation*}
Thus, we obtain four general solutions, which we write as
\begin{equation}
A^u =
C_1\exp\left(-\iota\sqrt{\frac{\omega^2}{(\alpha^u)^2}-\kappa^2}\,x_3\right) + 
C_2\exp\left(\iota\sqrt{\frac{\omega^2}{(\alpha^u)^2}-\kappa^2}\,x_3\right)\,,
\label{eq:QRGenSolAu}
\end{equation}
\begin{equation*}
\begin{aligned}
A^d &=
C_4\exp\left(-\sqrt{\kappa^2-\frac{\omega^2}{(\alpha^d)^2}}\,x_3\right)\,, \\
B^u &=
D_1\exp\left(-\iota\sqrt{\frac{\omega^2}{(\beta^u)^2}-\kappa^2}\,x_3\right) + 
D_2\exp\left(\iota\sqrt{\frac{\omega^2}{(\beta^u)^2}-\kappa^2}\,x_3\right)\,, \\
B^d &=
D_4\exp\left(-\sqrt{\kappa^2-\frac{\omega^2}{(\beta^d)^2}}\,x_3\right)\,.
\end{aligned}
\end{equation*}
Our assumption about the behaviour of solutions in the halfspace forces~$r^d$ and~$s^d$ to be real. Thus, we write the nonzero components of the displacement vector as
\begin{align}
\label{eq:QRuD1}
u_1^d=&\frac{\partial{\mathcal P}^d}{\partial x_1}-\frac{\partial {\mathcal S}^d}{\partial x_3}\\
\nonumber=&\left[\iota\kappa C_4\exp\left(-\kappa r^d x_3\right)
+D_4\kappa s^d\exp\left(-\kappa s^d x_3\right)\right]
\exp(\iota(\kappa x_1-\omega t))\,,
\end{align}
\begin{align}
u_3^d=&\frac{\partial{\mathcal P}^d}{\partial x_3}+\frac{\partial{\mathcal S}^d}{\partial x_1}\\
\nonumber=&\left[-C_4\kappa r^d\exp\left(-\kappa r^d x_3\right)
+\iota\kappa D_4\exp\left(-\kappa s^d x_3\right)\right]
\exp(\iota(\kappa x_1-\omega t))\,,
\end{align}
\begin{align}
\label{eq:QRuU1}
u_1^u=&\frac{\partial{\mathcal P}^u}{\partial x_1}-\frac{\partial{\mathcal S}^u}{\partial x_3}\\
\nonumber=&\left[\iota\kappa C_1 \exp(-\iota\kappa r^u x_3)+\iota\kappa C_2\exp(\iota\kappa r^u x_3)\right.\\
\nonumber &\left. \qquad+\,\iota\kappa s^u D_1\exp(-\iota\kappa s^u x_3)-\iota\kappa s^u D_2\exp(\iota\kappa s^u x_3)
\right]
\exp(\iota(\kappa x_1-\omega t))\,,
\end{align}
\begin{align}
\label{eq:QRuU3}
u_3^u=&\frac{\partial{\mathcal P}^u}{\partial x_3}+\frac{\partial{\mathcal S}^u}{\partial x_1}\\
\nonumber=&\left[-\iota\kappa r^u C_1 \exp(-\iota\kappa r^u x_3)+\iota\kappa r^u C_2\exp(\iota\kappa r^u x_3)\right.\\
\nonumber &\left. \qquad+\,\iota\kappa D_1\exp(-\iota\kappa s^u x_3)+\iota\kappa D_2\exp(\iota\kappa s^u x_3)
\right]
\exp(\iota(\kappa x_1-\omega t))\,,
\end{align}
which allows us to apply the boundary conditions.
%%%%%%%%%%%%%%%%%%%%%%%%%%%%
\subsection{Boundary conditions}
%%%%%%%%%%%%%%%%%%%%%%%%%%%%
Let us examine expressions~(\ref{eq:QRuD1})--(\ref{eq:QRuU3}) in view of Hooke's law, 
\begin{equation}
\label{eq:HookeLaw}
\sigma_{ij} =
\lambda\delta_{ij}\sum_{k=1}^3 \varepsilon_{kk} + 2\mu\varepsilon_{ij}
= (C_{11}-2C_{44})\,\delta_{ij}\sum_{k=1}^3\frac{\partial u_k}{\partial x_k} + 
C_{44}\left(\frac{\partial u_i}{\partial x_j}+\frac{\partial u_j}{\partial x_i}\right)\,,
\end{equation}
and the boundary conditions at~$x_3=0$\,, which are~$\sigma_{33}^u=\sigma_{31}^u=0$\,; hence, the first condition implies
\begin{equation*}
\left.\sigma_{31}^u\right|_{x_3=0}=0\Rightarrow \left.\frac{\partial u_1^u}{\partial x_3}\right|_{x_3=0}= -\left.\frac{\partial u_3^u}{\partial x_1}\right|_{x_3=0}\,.
\end{equation*}
Factoring out~$\exp(\iota(\kappa x_1-\omega t))$\,, we write
\begin{equation*}
\kappa^2 r^u C_1-\kappa^2 r^u C_2+\kappa^2 (s^u)^2D_1 +\kappa^2 (s^u)^2D_2
=-\kappa^2 r^u C_1+\kappa^2 r^u C_2+\kappa^2 D_1 +\kappa^2 D_2\,,
\end{equation*}
which, upon rearranging and factoring out~$\kappa^2$\,, we rewrite as
\begin{equation}
\label{eq:QRBC1}
2r^u(C_1-C_2)+[(s^u)^2-1](D_1+D_2)=0\,.
\end{equation}
The second condition implies
\begin{equation*}
\left.\sigma_{33}^u\right|_{x_3=0}=0\Rightarrow \left.\left[(C_{11}^u-2C_{44}^u)
\left(\frac{\partial u_1^u}{\partial x_1}+\frac{\partial u_3^u}{\partial x_3}\right)
+2C_{44}^u\left(\frac{\partial u_3^u}{\partial x_3}\right)\right]\right|_{x_3=0}=0\,,
\end{equation*}
which can be rearranged to
\begin{equation*}
 \left.\left[(C_{11}^u-2C_{44}^u)
\left(\frac{\partial u_1^u}{\partial x_1}\right)
+C_{11}^u\left(\frac{\partial u_3^u}{\partial x_3}\right)\right]\right|_{x_3=0}=0\,,
\end{equation*}
and, upon factoring out~$\kappa^2\exp(\iota(\kappa x_1-\omega t))$\,, further reduces to
\begin{equation}
(C_{11}^u-2C_{44}^u)\left[-(C_1+C_2)-s^u(D_1-D_2)\right]
+C_{11}^u\left[-(r^u)^2(C_1+C_2)+s^u(D_1-D_2)\right]=0\,.
\label{eq:sig330}
\end{equation}
At~$x_3=Z$\,, the boundary conditions are
\begin{align*}
u_1^u|_{x_3=Z} &= u_1^d|_{x_3=Z}\,,\quad u_3^u|_{x_3=Z}=u_3^d|_{x_3=Z}\,, \\
\sigma_{33}^u|_{x_3=Z} &= \sigma_{33}^d|_{x_3=Z}\,,\quad \sigma_{31}^u|_{x_3=Z}=\sigma_{31}^d|_{x_3=Z}\,.
\end{align*}
Factoring out ~$\kappa \exp(\iota(\kappa x_1-\omega t))$\,, the first condition becomes
\begin{align}
\iota C_1\exp(-\iota\kappa r^u Z)+\iota C_2\exp(\iota\kappa r^u Z) + \iota s^u D_1
& \exp(-\iota\kappa s^u Z)-\iota s^u D_2\exp(\iota\kappa s^u Z)\nonumber\\
=&\,\iota C_4\exp(-\kappa r^d Z)+s^d D_4\exp(-\kappa s^d Z)\,.
\label{eq:u1Z}
\end{align}
Similarly,  the second condition implies
\begin{align}
-\iota r^u C_1\exp(-\iota\kappa r^u Z)+\iota r^u C_2\exp(\iota\kappa r^u Z) + \iota 
&D_1 \exp(-\iota\kappa s^u Z)+\iota D_2\exp(\iota\kappa s^u Z)\nonumber\\
=&-r^d C_4\exp(-\kappa r^d Z)+\iota D_4\exp(-\kappa s^d Z)\,.
\label{eq:u3Z}
\end{align}
The third condition,
{\small
\begin{equation*}
 \left.\left[(C_{11}^u-2C_{44}^u)
\left(\frac{\partial u_1^u}{\partial x_1}\right)
+C_{11}^u\left(\frac{\partial u_3^u}{\partial x_3}\right)\right]\right|_{x_3=Z}
=
 \left.\left[(C_{11}^d-2C_{44}^d)
\left(\frac{\partial u_1^d}{\partial x_1}\right)
+C_{11}^d\left(\frac{\partial u_3^d}{\partial x_3}\right)\right]\right|_{x_3=Z}\,,
\end{equation*}}
upon factoring out~$\kappa^2\exp(\iota(\kappa x_1-\omega t))$\,, becomes
{%\small
\begin{align*}
(C_{11}^u&-2C_{44}^u)\left[-C_1\exp(-\iota\kappa r^u Z)
-C_2\exp(\iota\kappa r^u Z)
-s^u D_1\exp(-\iota\kappa s^u Z)
+s^u D_2\exp(\iota\kappa s^u Z)\right]\nonumber\\
+\,&C_{11}^u\left[-(r^u)^2 C_1\exp(-\iota\kappa r^u Z)
-(r^u)^2C_2\exp(\iota\kappa r^u Z)
+s^u D_1 \exp(-\iota\kappa s^u Z)
-s^u D_2\exp(\iota\kappa s^u Z)\right]\nonumber\\
&=
(C_{11}^d-2C_{44}^d)\left[
-C_4\exp(-\kappa r^d Z)
+\iota s^d D_4\exp(-\kappa s^d Z)\right]\nonumber\\
&\qquad+C_{11}^d\left[
(r^d)^2 C_4 \exp(-\kappa r^d Z)
-\iota s^d D_4 \exp(-\kappa s^d Z)\right]\,.
%\label{eq:sig33Z}
\end{align*}}
The fourth condition,
\begin{equation*}
\left. C_{44}^u\left(\frac{\partial u_1^u}{\partial x_3}+\frac{\partial u_3^u}{\partial x_1}\right)
\right|_{x_3=Z}=
\left. C_{44}^d\left(\frac{\partial u_1^d}{\partial x_3}+\frac{\partial u_3^d}{\partial x_1}\right)
\right|_{x_3=Z}\,,
\end{equation*}
implies
{%\small
\begin{align}
C_{44}^u&\left[
r^u C_1\exp(-\iota\kappa r^u Z)
-r^u C_2\exp(\iota\kappa r^u Z)
+(s^u)^2 D_1 \exp(-\iota\kappa s^u Z)
+(s^u)^2 D_2 \exp(\iota\kappa s^u Z)\right.\nonumber\\
&\left.+\,r^u C_1\exp(-\iota\kappa r^u Z)
-r^u C_2\exp(\iota\kappa r^u Z)
-D_1\exp(-\iota\kappa s^u Z)
-D_2\exp(\iota\kappa s^u Z)\right]\nonumber\\
&=
C_{44}^d\left[
-\iota r^d C_4\exp(-\kappa r^d Z)
-(s^d)^2D_4\exp(-\kappa s^d Z)
-\iota r^d C_4\exp(-\kappa r^d Z)
-D_4\exp(-\kappa s^d Z)\right]\,.
\label{eq:QRBC6}
\end{align}}
For a notational convenience, we let 
\begin{equation}
\label{eq:QRConvI}
a':=\kappa r^u Z\,, 
a=\kappa r^d Z\,, 
C_1':=C_1\,{\rm e}^{-\iota a'}\,,
C_2':=C_2\,{\rm e}^{\iota a'}\,, 
C_4':=C_4\,{\rm e}^{-a}\,,
\end{equation}
\begin{equation}
\label{eq:QRConvII}
b':=\kappa s^u Z\,,
b=\kappa s^d Z\,,
D_1':=D_1\,{\rm e}^{-\iota b'}\,,
D_2':=D_2\,{\rm e}^{\iota b'}\,,
D_4':=D_4\,{\rm e}^{-b}\,.
\end{equation}
Thus, conditions~(\ref{eq:QRBC1}) to~(\ref{eq:QRBC6}) can be written as
\begin{equation}
\label{eq:QRModBC1}
2r^u\,{\rm e}^{\iota a'}C_1'-2r^u\,{\rm e}^{-\iota a'}C_2'+[(s^u)^2-1]\,{\rm e}^{\iota b'}D_1'
+[(s^u)^2-1]\,{\rm e}^{-\iota b'}D_2'=0\,,
\end{equation}
\begin{align}
\label{eq:QRModBC2}
[-(C_{11}^u-&2C_{44}^u)-C_{11}^u (r^u)^2]\,{\rm e}^{\iota a'}C_1'
+[-(C_{11}^u-2C_{44}^u)-C_{11}^u (r^u)^2]\,{\rm e}^{-\iota a'}C_2'\nonumber\\
&+2C_{44}^u s^u\,{\rm e}^{\iota b'}D_1'-2C_{44}^u s^u\,{\rm e}^{-\iota b'}D_2'=0\,,
\end{align}
\begin{equation}
\label{eq:QRModBC3}
\iota C_1'+\iota C_2' +\iota s^u D_1' -\iota s^u D_2'-\iota C_4'-s^d D_4'=0\,,
\end{equation}
\begin{equation}
-\iota r^u C_1'+\iota r^u C_2'+\iota D_1'+\iota D_2'+r^d C_4'-\iota D_4'=0\,,
\label{eq:QRModBC4}
\end{equation}
\begin{align}
\label{eq:QRModBC5temp}
-(C_{11}^u&-2C_{44}^u)C_1'-(C_{11}^u-2C_{44}^u)C_2'-(C_{11}^u-2C_{44}^u)s^u D_1'
+(C_{11}^u-2C_{44}^u)s^u D_2'\nonumber\\
&-C_{11}^u (r^u)^2C_1'-C_{11}^u(r^u)^2C_2'+C_{11}^u s^u D_1'-C_{11}^u s^u D_2'
+(C_{11}^d-2C_{44}^d)C_4'
\nonumber\\&\qquad-\iota(C_{11}^d-2C_{44}^d)s^d D_4'
-C_{11}^d (r^d)^2 C_4'+\iota C_{11}^d s^d D_4'=0\,,
\end{align}
\begin{align}
\label{eq:QRModBC6temp}
C_{44}^u& r^u C_1' - C_{44}^u r^u C_2' +C_{44}^u (s^u)^2D_1'+C_{44}^u (s^u)^2 D_2'
+C_{44}^u r^u C_1'-C_{44}^u r^u C_2' -C_{44}^u D_1' -C_{44}^u D_2'\nonumber\\
&+\iota r^d C_{44}^d C_4'+C_{44}^d (s^d)^2 D_4'+\iota r^d C_{44}^d C_4'+C_{44}^d D_4'=0\,.
\end{align}
Simplifying, we write conditions~(\ref{eq:QRModBC5temp}) and (\ref{eq:QRModBC6temp}) as
\begin{align}
\label{eq:QRModBC5}
-[C_{11}^u&-2C_{44}^u+(r^u)^2C_{11}^u]C_1'
-[C_{11}^u-2C_{44}^u+(r^u)^2C_{11}^u]C_2'
+2C_{44}^u s^u D_1'-2C_{44}^u s^u D_2'\nonumber\\
&+[C_{11}^d-2C_{44}^d-(r^d)^2C_{11}^d]C_4'
+2\iota s^d C_{44}^d D_4'=0\,,
\end{align}
\begin{align}
\label{eq:QRModBC6}
2C_{44}^u r^u C_1'-2C_{44}^u r^u C_2'+C_{44}^u[(s^u)^2-1]D_1'&+C_{44}^u[(s^u)^2-1]D_2'
\nonumber\\
&+2\iota r^d C_{44}^d C_4'+C_{44}^d[(s^d)^2+1]D_4'=0\,.
\end{align}
%%%%%%%%%%%%%%%%%%%%%%%%%%%%%%%%%%
\subsection{Dispersion relation}
\label{sec:dispersion}
%%%%%%%%%%%%%%%%%%%%%%%%%%%%%%%%%%
The six boundary conditions stated in equations~(\ref{eq:QRModBC1}), (\ref{eq:QRModBC2}), (\ref{eq:QRModBC3}), (\ref{eq:QRModBC4}), (\ref{eq:QRModBC5}) and (\ref{eq:QRModBC6})
form a linear system of six equations for six unknowns,~$C_1'$\,,~$C_2'$\,,~$D_1'$\,,~$D_2'$\,,~$C_4'$ and~$D_4'$\,.
For a nontrivial solution, the determinant of the coefficient matrix,~$M_{r}$\,, must be zero.
Upon factoring~$\iota$ from the third and fourth rows, we write
{%
\begin{align}
M_r=&\left[
\begin{array}{ccc}
2 r^u\,{\rm e}^{\iota a'} &  -2 r^u\,{\rm e}^{-\iota a'} &  \dfrac{v_{r}^2\rho^u-2 C^u_{44}}{C^u_{44}}\,{\rm e}^{\iota b'}  \\ 
 \left[2 C_{44}^u-\rho^u v_{r}^2\right]\,{\rm e}^{\iota a'} &    \left[2 C_{44}^u-\rho^u v_{r}^2\right]\,{\rm e}^{-\iota a'} &  2C_{44}^us^u \,{\rm e}^{\iota b'}\\
1 &  1 &s^u \\
-r^u &    r^u &  1 \\
2 C_{44}^u-\rho^u v_{r}^2  &  2 C_{44}^u-\rho^u v_{r}^2 &  2C_{44}^u s^u \\
2C_{44}^{\mathstrut u} r^u &  -2C_{44}^u r^u &  v_{r}^2\rho^u-2 C^u_{44} 
\end{array}
\right.\nonumber\\
&\hspace*{1.5in}\left.\begin{array}{ccc}
\dfrac{v_{r}^2\rho^u-2 C^u_{44}}{C^u_{44}}\,{\rm e}^{-\iota b'} &  0 &  0\\
-2C_{44}^us^u \,{\rm e}^{-\iota b'} &  0 &  0\\
-s^u &-1 &  \iota s^d\\
1 &  -\iota r^d &  -1\\
-2C_{44}^u s^u &  v_{r}^2\rho^d-2 C^d_{44} &  2\iota C_{44}^d s^d\\
v_{r}^2\rho^u-2 C^u_{44} &  2\iota C_{44}^d r^d &   2 C^d_{44}-v_{r}^2\rho^d
\end{array}
\right]\,,
\label{eq:M_r}
\end{align}}

where we use
\begin{equation*}
C_{11}^u((r^u)^2+1)=C_{11}^u\frac{v_{r}^2}{(\alpha^u)^2}=\rho^u v_{r}^2\,\,,\quad (s^u)^2-1=\frac{v_{r}^2\rho^u-2 C^u_{44}}{C^u_{44}}\,,
\end{equation*}
\begin{equation*}
C^d_{11}(1-(r^d)^2)-2 C^d_{44}=v_{r}^2\rho^d-2 C^d_{44}\,\,,\quad C^d_{44}((s^d)^2+1)=2 C^d_{44}-v_{r}^2\rho^d\,.
\end{equation*}
%%%%%%%%%%%%%%%%%%%%%%%%%%%%%%%%
\section{Details of dispersion relation derivation}
\label{app:disprel}
%%%%%%%%%%%%%%%%%%%%%%%%%%%%%%%%
In this appendix, for the reader's convenience, we present operations to compute the determinant of the~$6\times 6$ matrix stated in expression~(\ref{eq:M_r}).
We invoke several algebraic properties that allow us to obtain a~$2\times 2$ matrix.
In this process, we use the following notational abbreviations.
\begin{itemize}
\item[$\circ$] $C1,\ldots,C6$ denotes columns~$1, \ldots,6$
\item[$\circ$] $R1,\ldots,R6$ denotes rows~$1, \ldots,6$
\item[$\circ$] $C1\mapsto C1+C2$ denotes replacement of~$C1$ by~$C1+C2$\,, etc. 
\item[$\circ$] $A':=\sin a'/r^u$
\item[$\circ$] $B':=\sin b'/s^u$
\item[$\circ$] $u_\beta:=v_{r}/\beta^u$
\end{itemize}
Using this notation, we perform the the following sequence of operations.\\
\begin{tabular}{p{6.5cm}p{6.5cm}}
\begin{enumerate}
%\addtolength{\itemsep}{-2pt}
	\item Factor out~$1/C^u_{44}$ from~$R1$
	\item $C1\mapsto C1-C2$
	\item Factor out 2 from~$C1$
	\item $C2\mapsto C2+C1$
	\item $C3\mapsto C3-C4$
	\item Factor out 2 from~$C3$
	\item $C4\mapsto C3+C4$
	\item Factor out~$r^u$ from~$C1$ and~$s^u$ from~$C3$
    \item $C2\mapsto C2- C3$
    \item $C4\mapsto C4+C1$   
\end{enumerate}
&
\begin{enumerate}
	\setcounter{enumi}{10}
	\item $R6\mapsto R6-2 C^u_{44} R4$
	\item $R5\mapsto R5-2 C^u_{44} R3$
	\item Factor out~$\iota$ from~$C2, C3, C5$
	\item Factor out~$-\iota$ from~$R3,R2, R5$
	\item Factor out~$C^u_{44}$ from~$ R1$  and~$R2$
	\item Move~$C3$ to the first column and shift the former first column and second column to the right; in other words, let~$C1'=C1$\,,~$C2'=C2$\,,~$C1\mapsto C3$\,,~$C2\mapsto C1'$\,,~$C3\mapsto C2'$\,.
	\end{enumerate}
\end{tabular}
{\small
\begin{align*}
\det[M_r]= 4r^u s^u  C^u_{44}\det&\left[
\begin{array}{ccc}
((u_\beta)^2-2) B' & 2 \cos a' & 2 (r^u)^2 A' -((u_\beta)^2-2)B'\\
 2 \cos b'  &  ((u_\beta)^2-2) A'  & - 2 \cos b'-((u_\beta)^2-2)\cos a'\\
1 & 0 & 0 \\
0 &   -1 & 0\\
0  & 0 &  - v_{r}^2\rho^u\\
0  & 0 & 0
\end{array}
\right.\\
&\left.\begin{array}{ccc}
((u_\beta)^2-2)\cos b'+2 \cos a' & 0 &0\\ 
((u_\beta)^2-2)A' -2 (s^u)^2B' & 0 &0\\
0 & -1 &-s^d\\
0  & - r^d &-1\\
0 &v_{r}^2\rho^d-2  (C^d_{44}-C^u_{44}) &  - 2 s^d(C^d_{44}-C^u_{44})\\
v_{r}^2\rho^u & 2 r^d(C^d_{44}-C^u_{44}) &2 (C^d_{44}-C^u_{44})-v_{r}^2\rho^d
\end{array}\right]\,.
\end{align*}}
Let us write the above matrix in the block form, 
\begin{equation*}
\left[\begin{array}{cc} B & 0 \\ A & C\end{array}\right]\,,
\end{equation*}
where
{\small
\begin{equation*}
B:=\left[
\begin{array}{cccc}
((u_\beta)^2-2) B' &2 \cos a' &2 (r^u)^2 A' -((u_\beta)^2-2)B' &((u_\beta)^2-2)\cos b'+2 \cos a'\\ 
 2 \cos b'  & ((u_\beta)^2-2) A'  &- 2 \cos b'-((u_\beta)^2-2)\cos a'  &((u_\beta)^2-2)A' -2 (s^u)^2B'
\end{array}\right]\,,
\end{equation*}}
\begin{equation*}
A:=\left[\begin{array}{cccc}1 & 0 & 0 & 0  \\
0 &   -1 & 0 & 0 \\
0  & 0 &  - v_{r}^2\rho^u  & 0 \\
0  & 0 & 0 & v_{r}^2\rho^u   
\end{array}\right]\!,
\quad
C:=\left[\begin{array}{cc}
-1 & - s^d\\
 - r^d & - 1\\
v_{r}^2\rho^d-2  (C^d_{44}-C^u_{44}) & - 2 s^d(C^d_{44}-C^u_{44}) \\
 2 r^d(C^d_{44}-C^u_{44})   & 2 (C^d_{44}-C^u_{44})-v_{r}^2\rho^d 
\end{array}\right].
\end{equation*}
Since~$A$ is invertible, we have
{%\small
\begin{equation*}
\left[ \begin{array}{cc} B & 0 \\ A & C\end{array}\right]=\left[ \begin{array}{cc} 0 & I_2 \\ I_4 & 0\end{array}\right] \left[ \begin{array}{cc} A & C \\ B & 0\end{array}\right]=
\left[ \begin{array}{cc} 0 & I_2 \\ I_4 & 0\end{array}\right] \left[ \begin{array}{cc} A & 0 \\ 0 & I_2\end{array}\right] \left[ \begin{array}{cc} I_4 & 0 \\B  & -I_2\end{array}\right]
\left[ \begin{array}{cc} I_4 & A^{-1} C  \\ 0 & BA^{-1} C\end{array}\right]\,,
\end{equation*}}
and, hence,
\begin{equation*}
\det\left[ \begin{array}{cc} B & 0 \\ A & C\end{array}\right]=\det[A]\det[B A^{-1}C]\det \left[ \begin{array}{cc} 0 & I_2 \\ I_4 & 0\end{array}\right] 
\det \left[ \begin{array}{cc} I_4 & 0 \\ B  & -I_2\end{array}\right]=\det[A]\det[B A^{-1}C].
\end{equation*}
Finally, we can write the determinant of the coefficient matrix as
\begin{equation*}
\det[M_r]=4r^u s^u  C^u_{44}(XY -ST)=4C^u_{44} \det\left[\begin{array}{cc} s^u X & s^u S\\r^u T & r^u Y\end{array}\right]\,,
\end{equation*}
where~$X, Y, S, T$ are given equation~(\ref{eq:ourdet}).
%%%%%%%%%%%%%%%%%%%%%%%%%%%%
\section{Comparison to results in literature}
\label{app:comparison}
%%%%%%%%%%%%%%%%%%%%%%%%%%%%
In this appendix, we present the notational translations for the quasi-Rayleigh waves in the literature, and compare differences found among them.
Herein, for convenience, we use definitions~(\ref{eq:QRConvI}) and (\ref{eq:QRConvII}).

Love~\cite{Love1911} considered the same problem but with the assumption of incompressibility.
Lee~\cite{Lee1932}, Fu~\cite{Fu1946} and Ud\'{i}as~\cite{Udias1999} considered the same problem without the assumption of incompressibility but made other simplifying assumptions.
%%%%%%%%%%%%%%%%%%%%%%%%%%%%%%%%
\subsection{Comparison with Love~\cite{Love1911}}
%%%%%%%%%%%%%%%%%%%%%%%%%%%%%%%%
Love~\cite{Love1911} simplifies his formulation by assuming incompressibility, which implies that~$\alpha^u\rightarrow\infty$ and~$\alpha^d\rightarrow\infty$\,.
As a consequence,~$r^u=\iota$ and~$r^d=1$\,. 
It follows that
\begin{equation*}
\cos(a')=\cos(\kappa r^uZ)=\cos(\iota\kappa Z)=\cosh(\kappa Z)\,,
\end{equation*}
\begin{equation*}
\sin(a')=\sin(\kappa r^uZ)=\sin(\iota\kappa Z)=\iota\sinh(\kappa Z)\,,
\end{equation*}
\begin{equation*}
\cosh(\iota\kappa s^uZ)=\cosh(\iota b')=\cos(b')\,,
\end{equation*}
\begin{equation*}
\sinh(\iota\kappa s^uZ)=\sinh(\iota b')=\iota\sin(b')\,.
\end{equation*}
Consequently, his
\begin{equation*}
\xi\eta'-\xi'\eta=0
\end{equation*}
becomes
\begin{equation*}
\frac{\iota s^u}{(C_{44}^u)^2}(XY-ST)=\frac{\iota s^u}{(C_{44}^u)^2}(\mathrm{D}_{r2})=\frac{\iota\,\mathrm{D}_r}{4(C_{44}^u)^3\,r^u}=0\,,
\end{equation*}
in our notation, where~$\mathrm{D}_r$ is defined by expression~(\ref{eq:ourdet}).
Under the assumption of incompressibility,~$r^u$ becomes~$\iota$\,, and thus
\begin{equation*}
\xi\eta'-\xi'\eta=\frac{\mathrm{D}_r}{4(C_{44}^u)^3}\,.
\end{equation*}
%%%%%%%%%%%%%%%%%%%%%%%%%%%%%%%%
\subsection{Comparison with Lee~\cite{Lee1932}}
%%%%%%%%%%%%%%%%%%%%%%%%%%%%%%%%
Lee~\cite{Lee1932} obtains a determinantal equation in terms of trigonometric functions.
In our notation, his determinantal equation,
\begin{equation*}
\xi\eta'-\xi'\eta=0\,,
\end{equation*}
becomes
\begin{equation*}
\frac{XY-ST}{(C_{44}^u)^2}=\frac{\mathrm{D}_{r2}}{(C_{44}^u)^2}=\frac{\mathrm{D}_r}{4r^us^u(C_{44}^u)^3}=0\,.
\end{equation*}
In other words, our expressions differ by a multiplicative factor.
%%%%%%%%%%%%%%%%%%%%%%%%%%%%%%%%
\subsection{Comparison with Fu~\cite{Fu1946}}
%%%%%%%%%%%%%%%%%%%%%%%%%%%%%%%%
Fu~\cite{Fu1946} obtains a determinantal equation in terms of hyperbolic functions.
Invoking standard expressions,
\begin{equation*}
\cosh(\iota a')=\cos(a')\,,\quad\cosh(\iota b')=\cos(b')\,,\quad
\sinh(\iota a')=\iota\sin(a')\,,\quad\sinh(\iota b')=\iota\sin(b')\,,
\end{equation*}
we write his determinantal equation,
\begin{equation*}
\zeta\eta'-\zeta'\eta=0\,,
\end{equation*}
as
\begin{equation*}
(XY-ST)\,\kappa^8=\mathrm{D}_{r2}\,\kappa^8=\frac{\kappa^8\,\mathrm{D}_r}{4\,C_{44}^u\,r^u\,s^u}=0\,,
\end{equation*}
in our notation.
%%%%%%%%%%%%%%%%%%%%%%%%%%%%%%%%
\subsection{Comparison with Ud\'{i}as~\cite{Udias1999}}
%%%%%%%%%%%%%%%%%%%%%%%%%%%%%%%%
\label{sec:udias}
Referring to Ud\'{i}as~\cite{Udias1999}, his determinantal equation,
\begin{equation*}
\xi\eta'-\xi'\eta=0\,,
\end{equation*}
becomes
\begin{equation*}
\frac{XY-ST}{(C_{44}^u)^2}=\frac{\mathrm{D}_{r2}}{(C_{44}^u)^2}=\frac{\mathrm{D}_r}{4\,(C_{44}^u)^3\,r^u\,s^u}=0\,,
\end{equation*}
where we use the following six corrections to Ud\'{i}as's formu{l\ae}.
\begin{enumerate}
\item
The second term of formula~(10.87),~$Zr'\sin a'$\,, should be~$(Z/r')\sin a'$\,.
\item
In the third term of formula~(10.87),~$\sin a'$ should be~$\sin b'$\,.
\item
In the fourth term of formula~(10.86),~$\sin b'$ should be~$\cos b'$\,.
\item
The~$(\beta')^2$ in the first denominator of formula~(10.92) should be~$\beta^2$\,.
\item
Instead of~$r$ and~$s$\,, Ud\'{i}as~\cite{Udias1999} should have~$\overline{r}$ and~$\overline{s}$\,, which are the magnitudes of~$r$~and~$s$\,.
\item
Our determinant, with the above corrections to Ud\'{i}as~\cite{Udias1999}, is~$4(C_{44}^u)^3 r^u s^u$ times his formula~(10.85); thus, formula~(10.85)  does not include solutions~$r^u=0$ and~$s^u=0$\,.
However, as we discuss in Section~\ref{sec:quasirayl}, those solutions exhibit zero displacements, which might be the reason why Lee~\cite{Lee1932}, Fu~\cite{Fu1946}, Ud\'{i}as~\cite{Udias1999} and Ben-Menahem and Singh~\cite[Section~3.6.5]{benmenahem}
omit them.
\end{enumerate}
%%%%%%%%%%%%%%%%%%%%%%%%%%%%%%%%
\subsection{Comparison with Ben-Menahem and Singh~\cite{benmenahem}.}
%%%%%%%%%%%%%%%%%%%%%%%%%%%%%%%%
Translation of the notation of
Ben-Menahem and Singh~\cite[Section~3.6.5]{benmenahem}
into our notation results in
\begin{equation}
\Delta_R=\frac{-\kappa^8}{(C_{44}^u)^2}(XY-ST)=\frac{-\kappa^8\mathrm{D}_{r2}}{(C_{44}^u)^2}=
\frac{-\kappa^8\mathrm{D}_r}{4(C_{44}^u)^3 r^u s^u}=0\,.
\end{equation}
%%%%%%%%%%%%%%%%%%%%%%%%%%%%%%%%
\end{document}